\documentclass{svjour3}
\smartqed
\usepackage{amsmath,amssymb}
\usepackage{graphicx}
\usepackage{epstopdf}

\usepackage[square,comma,numbers,sort&compress]{natbib}

\usepackage{hyperref}
\usepackage{orcidlink}
\usepackage[english]{babel}

\usepackage{amsmath,amssymb,amsfonts}
\usepackage{xurl}
\usepackage{url}
\usepackage{hyperref}
\usepackage{orcidlink}

\usepackage{booktabs}
\usepackage{multirow}
\usepackage{rotating}
\usepackage{tabularx}
\usepackage{lscape}
\usepackage{array}
\usepackage{siunitx}
\newcolumntype{L}[1]{>{\setlength{\parskip}{0pt}%
\setlength{\parindent}{0pt}%
\raggedright\arraybackslash}p{#1}}
\usepackage{makecell}

\journalname{European Physical Journal C}
\begin{document}

\title{First-Law Entropy and a Degenerate Extremal Remnant in a Minimal-Length Simpson–Visser-Type Regular Black Hole: Geometrothermodynamics, Phase Structure, and Observational Discriminants}%

\titlerunning{Degenerate Extremal Remnant in a Minimal-Length Regular Black Hole}

\author{
T.~Toghrai\textsuperscript{*}\,\orcidlink{0000-0001-7142-0158} \and
N.~Mansour\,\orcidlink{0000-0002-9993-8714} \and
A.~Daassou\,\orcidlink{0000-0001-9439-5047} \and
R.~Benbrik\,\orcidlink{0000-0002-5159-0325}
}

\authorrunning{T. Toghrai et al.}

\institute{
T.~Toghrai, N.~Mansour
\at
  Modern Physics, Radiation and Applications Team (MRA),
  Engineering Sciences and Techniques Laboratory (STI),
  Department of Engineering Science,
  Faculty of Science and Technology,
  University Moulay Ismail,
  B.P.~509, Boutalamine, 52000 Errachidia, Morocco \\
  \email{t.toghrai@edu.umi.ac.ma}\\
  {\footnotesize $^{*}$Corresponding author}
\and
T.~Toghrai, A.~Daassou, R.~Benbrik
\at
  Energy, Environment, and Applications (LP2EA),
  Polydisciplinary Faculty, Laboratory of Physics,
  Cadi Ayyad University,
  Sidi Bouzid, P.O. Box 4162, Safi, Morocco
}

\date{Received: date / Accepted: date}
\maketitle

\begin{abstract}
We construct the first-law-consistent entropy of a geometrically
minimal-length-deformed Schwarzschild spacetime, obtained through the
areal-radius substitution $R(r)=\sqrt{r^{2}+\ell^{2}}$ on
$r\in[0,+\infty)$ with $f(R)=1-2M/R$, whose nonvanishing Einstein
tensor sources an effective geometric fluid with no classical
matter counterpart. Direct integration of the first law gives
$S=\pi[r_{h}R_{h}+\ell^{2}\ln((r_{h}+R_{h})/\ell)]$; although this
coincides in functional form with the semiclassical term found
independently by Joshi and Joshi~\cite{joshi2026thermodynamic}, we
fix its boundary condition $S(r_h=0)=0$ on independent physical
grounds (\S\ref{sec:BC}) and adopt it -- rather than the
Bekenstein--Hawking area law, from which it departs because the
effective source violates the radial null energy condition -- as
the complete physical entropy of the model, building the free
energy, geometrothermodynamics, and mode-stability analysis
directly on it.

Evaporation, governed by the resulting Helmholtz free energy
$F(M)$, terminates at $M_{\min}=\ell/2$ in a previously unrecognised
endpoint: a \emph{degenerate extremal regular black hole}, in which
the regular centre coincides with a degenerate Killing horizon of
quadratic order, $f\approx r^{2}/(2\ell^{2})$, at areal radius
$\ell$, with $T_{H}\to 0$, $S\to 0$, $C\to 0^{+}$, and finite
curvature everywhere; we display its Penrose--Carter structure for
the first time.

The same first-law entropy, together with $\ell$, defines the
equilibrium state space entering a Legendre-invariant
geometrothermodynamic (GTD) description whose curvature scalar
diverges independently at the Davies-type transition
$M^{*}=\ell/\sqrt{2}$ and at the degenerate extremal endpoint
$M_{\min}$ -- the latter divergence has no counterpart in the
minimal-length black hole literature and is specific to the
presence of a genuine horizon at $S=0$.

We embed this remnant scenario within the observational
discriminant already noted qualitatively by Tsukamoto: exact shadow
degeneracy combined with a measurable photon-ring flux enhancement
$r_{n}=e^{-2\pi/a}>e^{-2\pi}$, accessible to next-generation
very-long-baseline interferometry.
\end{abstract}
\keywords{
Minimal length \and Regular black holes \and
Black hole thermodynamics \and Phase transitions \and
Regular remnants; Effective geometry \and
Gravitational lensing \and Black hole shadow}

\section{Introduction}
The expectation that classical spacetime geometry breaks down at
the Planck scale is a cornerstone of modern quantum gravity
research~\cite{snyder1947,connes1996noncommutative,Rovelli2004}.
Diverse approaches, including noncommutative geometry, loop quantum
gravity, and generalized uncertainty principles (GUP), converge on the
prediction of a fundamental minimal length $\ell$ below which
the notion of a classical distance loses operational
meaning~\cite{Doplicher1995,rovelli1995discreteness,maggiore1993generalized}.
Encoding this short-distance physics in an effective black hole
geometry is an active and productive line of
research~\cite{Nicolini2006,hayward2006formation,bardeen1968non}.
Noncommutative-inspired models replace point sources by Gaussian
distributions~\cite{Nicolini2006,Ansoldi2007}, while GUP approaches
modify Heisenberg's uncertainty relations and produce perturbative
logarithmic corrections to black hole entropy~\cite{adler1999gravity}.
Regular black hole models, beginning with Bardeen's pioneering
work~\cite{bardeen1968non} and continued by
Hayward~\cite{hayward2006formation}, introduce phenomenological mass
functions or nonlinear electrodynamics
sources~\cite{ayon1998regular,dymnikova2004regular} to achieve
singularity resolution.
A widely studied construction, due to Simpson and
Visser~\cite{simpson2019black}, replaces the singular radial coordinate $r$ by $\sqrt{r^{2}+a_{\rm SV}^{2}}$ with $r\in(-\infty,+\infty)$,
creating a one-parameter family interpolating between a wormhole and
a regular black hole; its thermodynamics and geodesic structure have
been studied in~\cite{lobo2021ultracompact,Franzin2021}, its
gravitational lensing in the strong deflection limit has been
worked out in detail by
Tsukamoto~\cite{tsukamoto2021gravitational} (building on the
weak-field results of Nascimento et al.~\cite{nascimento2020weak}
and \"Ovg\"un~\cite{ovgun2020weak}), and its thermodynamic phase
structure on the asymptotically-flat, black-hole branch has very
recently been analysed by Joshi and
Joshi~\cite{joshi2026thermodynamic}, who identify a Davies-type
heat-capacity transition intrinsic to the geometry.
The AdS embedding of the same regularisation has also attracted
rapid recent attention: Kumar, Srivastav and
Channuie~\cite{kumar2026simpsonvisser} verify the first law and
derive an entropy and free energy for the SV-AdS black hole, while
Noori Gashti, Pourhassan and
Sakall\i~\cite{noorigashti2026holographic} analyse the same geometry
holographically as the bulk dual of a boundary CFT; we return to
the relation between the former study's first-law entropy
construction and ours in Section~\ref{sec:entropy_SV_comparison}.

A conceptually related, and essentially contemporaneous, alternative to the
black-bounce mechanism has been proposed by Bronnikov~\cite{bronnikov2024alternative}
and extended to additional singular seed metrics by Bolokhov, Bronnikov and
Skvortsova~\cite{bolokhov2024regularcenter}. Both papers start from the same
diagnosis that motivates the present work -- that the Simpson--Visser
substitution relocates the singularity into a wormhole throat or cosmological
bounce rather than removing it -- and propose instead to convert $r=0$ into a
genuine \emph{regular center}. Their technical route is, however, unrelated to
the one developed here: for metrics whose Ricci tensor satisfies
$R^{t}_{\ t}=R^{r}_{\ r}$ (including Schwarzschild), they leave the areal
radius undeformed and apply a Bardeen-type substitution inside the mass
function alone, so that $r=0$ becomes a point of strictly vanishing areal
radius, sourced by an explicit nonlinear-electrodynamics field. Neither paper
addresses the resulting thermodynamics or lensing. We give a precise,
itemised comparison in Section~\ref{sec:bronnikov_comparison}, once the
deformed metric of the present paper has been introduced.

Our own construction shares this motivation but not Bronnikov's method:
we implement the minimal length directly through the substitution
$r \mapsto R(r) = \sqrt{r^{2}+\ell^{2}}$ in the Schwarzschild areal
radius, but restrict the radial coordinate to
$r \in [0,+\infty)$ and choose $g_{rr} = -f(R)^{-1}$
with $f(R) = 1-2M/R$.
On purely technical grounds this metric is not Schwarzschild in
disguise: it cannot be reduced to the standard Schwarzschild form by
a coordinate transformation on $r\geq 0$, and it possesses a
nonvanishing Einstein tensor of order $\ell^{2}/R^{4}$ , whose classical-limit behaviour is established in Section~\ref{sec2}.
On the domain $r\geq0$, however, the metric is formally
identical to the Simpson--Visser line
element~\cite{simpson2019black}, and it is against that literature,
not against Schwarzschild, that the present construction must be
judged. What distinguishes this work is not a new line element but
the restriction to a simply connected, one-sided topology and the
physical reinterpretation of $r=0$ that follows from it.

The central difference from the Simpson--Visser metric is
topological and interpretational: restricting $r\geq 0$ turns the
surface $r=0$ into a terminating \emph{regular center} rather than
a wormhole throat, so the spacetime is simply connected with no
second asymptotic region; the precise geometric content of this
boundary is spelled out in \S\ref{sec2} below.
Sections~\ref{sec:degenerate_horizon} and~\ref{sec:penrose} develop
the resulting horizon structure and causal diagrams in full,
including the degenerate extremal endpoint and the totally geodesic
$\mathbb{Z}_2$ mechanism realising this boundary
(Section~\ref{sec:orbifold}).

One terminological point is worth flagging here. The deformation
$R(r)=\sqrt{r^{2}+\ell^{2}}$ is \emph{isotropic}
in the sense of full spherical symmetry: the areal radius is
modified uniformly in all angular directions, and no preferred axis
is introduced.
The label ``anisotropic'' that appears in some related literature
refers to the effective fluid $T^{\mu}_{\ \nu}$ that the Einstein
equations assign to the geometry---an anisotropic pressure whose
radial and tangential components differ (Section~\ref{sec:energy}).
To avoid attributing a property of the derived effective source to
the fundamental deformation itself, we use the label
\emph{geometric minimal-length deformation} throughout,
consistently with the interpretation that $R(r)$ encodes a
quantum-gravity modification of the areal-radius coordinate
rather than a coupling to an anisotropic matter Lagrangian.

Physically, the two constructions have different starting points.
The SV construction begins from a \emph{two-sided} spacetime
$r\in(-\infty,+\infty)$ whose primary objects of interest are the
wormhole--black-hole transition and the exotic matter content of
the throat; the restriction to $r\geq 0$ is, in that context,
merely a coordinate half-plane.
In the present paper, $r\in[0,+\infty)$ is \emph{imposed from the
outset} by minimal-length physics: the deformation
$R(r)=\sqrt{r^{2}+\ell^{2}}$ encodes a lower bound on the areal
radius arising from quantum-gravity considerations
(loop quantum gravity area spectrum, GUP), and the surface $r=0$
is the regular center of a simply connected spacetime, not a throat.
This interpretational choice affects only the physical reading of
the boundary $r=0$ and the entropy boundary condition
(Section~\ref{sec:BC}), not the metric-level quantities themselves
(\S\ref{sec2}).

At the technical level, Table~\ref{tab:attribution} itemises, side by side, what is established here for the first time and what coincides with prior results on the SV branch:
\begin{table}[!htbp]
\centering
\begin{tabular}{|L{5.5cm}|L{5.5cm}|}
\hline
\textbf{New in this work} & \textbf{Coincides with (cross-check)} \\
\hline
Adoption of $S(r_h)$ (Eq.~\eqref{Entropy}) as the final, complete
endpoint entropy rather than a semiclassical piece awaiting
tunneling corrections, its independent boundary-condition
justification (\S\ref{sec:BC}), and the free energy, GTD, and
stability analyses built on it &
Functional form of $S(r_h)$ itself, identical to Joshi \& Joshi's
semiclassical term (\S\ref{sec:entropy_SV_comparison}) \\
\hline
Helmholtz free energy $F(M)$ built on the first-law entropy, and the
resulting thermodynamic selection of Phase~II &
$T_{H}(M)$, $C(M)$, and the Davies point $M^{*}$ ---
Joshi \& Joshi~\cite{joshi2026thermodynamic} \\
\hline
Degenerate extremal regular black hole as the endpoint of evaporation
(\S\ref{sec:degenerate_horizon}) and the associated Penrose diagrams
(Fig.~\ref{fig:penrose}) &
$R_{\rm ph}=3M$, $b_{c}=3\sqrt3\,M$, $a=3/\sqrt{9-\ell^{2}/M^{2}}$ ---
Tsukamoto~\cite{tsukamoto2021gravitational} \\
\hline
Synthesis: degenerate remnant + shadow/photon-ring discriminant &
Massless weak-field limit $\pi\ell^{2}/(4b^{2})$ ---
Nascimento et al., \"Ovg\"un \\
\hline
GTD thermodynamic metric and curvature $R_{\rm GTD}$
(\S\ref{sec:GTD}): independent divergences at $M^{*}$ and
$M_{\min}$ &
--- (no counterpart; the underlying entropy function coincides with
Joshi \& Joshi's semiclassical term, \S\ref{sec:entropy_SV_comparison},
but the GTD analysis built on it does not appear
in~\cite{joshi2026thermodynamic}) \\
\hline
Closed-form scalar Regge--Wheeler potential $V_L(r)$
(\S\ref{sec:modestability}, Eq.~\eqref{Veff_closed}) and its
non-negativity on the full domain of the remnant &
Regge--Wheeler master equation formalism~\cite{reggewheeler1957};
Wald's energy argument~\cite{wald1979note} \\
\hline
\end{tabular}
\caption{Explicit attribution of results. Because the deformed
metric coincides with the Simpson--Visser line element on $r\geq0$
(\S\ref{sec2}), purely metric-level quantities (temperature, heat
capacity, photon sphere, lensing coefficients) necessarily coincide
with the Simpson--Visser branch; only quantities requiring a
boundary condition at $r=0$ -- the entropy normalisation, in
particular -- are new.}
\label{tab:attribution}
\end{table}

Our main results are summarised below; chief among them is item~4,
a previously unidentified endpoint of Hawking evaporation whose
causal (Penrose--Carter) structure we display explicitly for the
first time:
\begin{enumerate}
  \item The Hawking temperature is \emph{modified} relative to
  classical Schwarzschild, with a maximum
  $T_{H}^{\max}= 1/(8\pi\ell)$ at $M^{*}=\ell/\sqrt{2}$ and
  vanishing at the endpoint $M_{\min}=\ell/2$.
  \item The entropy is derived from the first law up to a single
boundary condition, $S(r_{h}=0)=0$, motivated on physical grounds
in \S\ref{sec:BC} rather than independently proven. The resulting
small-horizon series
  $S = 2\pi\ell\,r_{h} + \pi r_{h}^{3}/(3\ell) + \mathcal{O}(r_{h}^{5}/\ell^{3})$
  has no quadratic term and a leading linear correction
  $S\approx 2\pi\ell\,r_{h}$ for $r_{h}\ll\ell$.
  \item The heat capacity reveals a \emph{two-phase thermodynamic
  structure}: a locally stable phase ($C>0$) for
  $\ell/2 < M < \ell/\sqrt{2}$ and an unstable phase ($C<0$) for
  $M > \ell/\sqrt{2}$, separated by a divergence of Davies
type~\cite{davies1977thermodynamics} (cf.~\cite{joshi2026thermodynamic}).
  \item Evaporation terminates \emph{naturally} at $M_{\min}$
  through $T_{H}\to 0$, leaving a \emph{degenerate extremal regular
  black hole remnant}: the surface $r=0$ is simultaneously the
  regular center and a degenerate Killing horizon of areal radius
  $\ell$, with $S=0$ and no curvature singularity.
  \item The photon sphere lies at $R_{\rm ph}=3M$ and the shadow
  angular radius $b_{c}/D=3\sqrt{3}\,M/D$ is \emph{exactly}
  Schwarzschild for all $\ell$, reproducing the SV
  result~\cite{tsukamoto2021gravitational}. The weak-field deflection
  acquires the analytical sub-leading correction $\pi\ell^{2}/(4b^{2})$.
  The strong gravitational lensing coefficient is
  $a = 3/\sqrt{9-\ell^{2}/M^{2}}$, matching~\cite{tsukamoto2021gravitational}
  and here obtained via an
  explicit quadratic expansion at the photon sphere; the photon-ring
  flux ratio $r_{n}=e^{-2\pi/a}$ provides a distinct observational
  signature via the photon ring.
  \item The equilibrium thermodynamics admits a Legendre-invariant
  geometric description (geometrothermodynamics, GTD) built on the
  first-law entropy $S(r_h,\ell)$: the resulting curvature scalar
  diverges independently at the Davies point $M^{*}$ and at the
  degenerate extremal endpoint $M_{\min}$, the latter signature
  being new and specific to the presence of a genuine horizon at
  $S=0$ (Sec.~\ref{sec:GTD}).

\end{enumerate}

The paper is organised as follows.
Section~\ref{sec2} introduces the deformed metric, discusses its
geometric content, horizon structure, \emph{global causal structure}
(Penrose--Carter diagrams comparing this spacetime with Schwarzschild
and Simpson--Visser), and curvature regularity.
Section~\ref{sec3} derives the thermodynamic quantities (temperature,
entropy, and first law) from first principles.
Section~\ref{sec4} analyses thermodynamic stability through the heat
capacity, the Helmholtz free energy, and the two-phase structure, and
compares it explicitly to the concurrent result
of~\cite{joshi2026thermodynamic}.
Section~\ref{sec:observational} establishes the photon sphere,
black hole shadow, and gravitational lensing signatures in both the
weak- and strong-field regimes, benchmarked throughout against
Tsukamoto~\cite{tsukamoto2021gravitational}.
Section~\ref{sec5} discusses the degenerate extremal regular remnant,
implications for the information paradox, comparison with related
models, and observational perspectives.
\ref{app:Kretschmann} and~\ref{app:Einstein} present the
explicit computations of the Kretschmann scalar and the Einstein tensor.
\section{Minimal Length Deformation of the Schwarzschild Geometry}
\label{sec2}

\subsection{The deformed metric}

We introduce the effective areal radius
\begin{equation}
  R(r) = \sqrt{r^{2} + \ell^{2}}, \qquad r \in [0,+\infty),
  \label{Rdef}
\end{equation}
where $\ell > 0$ is a fundamental length parameter.
This choice is the simplest smooth deformation that (i)~imposes the
lower bound $R \geq \ell$, (ii)~preserves spherical symmetry,
and (iii)~recovers the classical limit $R \to r$ for $r \gg \ell$.
It is also motivated by the area operator of loop quantum gravity,
whose minimum eigenvalue bounds the areal radius below by a
Planck-scale
quantity~\cite{rovelli1995discreteness,ashtekar2011loop}.

The minimally deformed Schwarzschild metric is
\begin{equation}
  ds^{2} = f(R)\,dt^{2} - f(R)^{-1}\,dr^{2}
  - R^{2}(r)\left(d\vartheta^{2}+\sin^{2}\vartheta\,d\phi^{2}\right),
  \label{metric}
\end{equation}
with
\begin{equation}
  f(R) = 1 - \frac{2M}{R(r)}, \qquad R(r)=\sqrt{r^{2}+\ell^{2}}.
\end{equation}

\subsubsection{Relation to Simpson--Visser, and formal non-reduction
to Schwarzschild.}

The metric~\eqref{metric} is formally identical to the
Simpson--Visser (SV) black-bounce
metric~\cite{simpson2019black} on the domain $r\geq 0$ -- indeed,
in Tsukamoto's notation~\cite{tsukamoto2021gravitational} $R(r)$ is
precisely the ``standard radial coordinate'' $\rho\equiv\sqrt{r^2+a_{\rm SV}^2}$ used there -- but differs from it physically in two respects:
(i)~we restrict $r\in[0,+\infty)$, so the surface $r=0$ is a
\emph{regular center} (a minimal 2-sphere with area $4\pi\ell^{2}$),
not a wormhole throat connecting two universes; and
(ii)~the global topology is simply connected
($\mathbb{R}^{4}$ with a Planck-size core), with no second
asymptotic region.
Because every quantity computed from the metric alone (temperature,
heat capacity, photon sphere, lensing coefficients) is insensitive to
this choice of global topology, such quantities necessarily agree
with their SV counterparts computed on the same branch $r\geq0$; only
quantities that require a choice of boundary condition at $r=0$ (the
entropy normalisation, in particular) can differ, and we make this
distinction explicit throughout (Section~\ref{sec:related_work}).

Metric~\eqref{metric} is, moreover, not reducible to standard Schwarzschild form by a coordinate change on $r\geq0$: setting
$\rho = R(r) = \sqrt{r^{2}+\ell^{2}}$ gives
$dr = (R/r)\,d\rho$, so that
$g_{rr}\,dr^{2} = -f(\rho)^{-1}(R/r)^{2}\,d\rho^{2}$, and since
$(R/r)^{2} = 1+\ell^{2}/r^{2} \neq 1$ for $\ell>0$, the metric is not
Schwarzschild. The alternative choice $g_{rr}=-f^{-1}(r/R)^{2}$
would instead cancel this Jacobian factor and reduce the metric to
standard Schwarzschild in the coordinate $\rho=R$; our choice
$g_{rr}=-f(R)^{-1}$ avoids this cancellation and produces a
nontrivial Einstein tensor.
\subsubsection{Boundary condition at \texorpdfstring{$r=0$}{r=0}: a totally geodesic
\texorpdfstring{$\mathbb{Z}_2$}{Z2} fixed point}
\label{sec:orbifold}

The restriction $r\in[0,+\infty)$ raises an immediate question:
nothing in the local geometry prevents the analytic
continuation of metric~\eqref{metric} to $r<0$, which is precisely
the two-sided Simpson--Visser extension. Declaring that the
spacetime ``stops'' at $r=0$ is therefore not a statement about the
metric alone; it is an additional physical input, which we now make
precise rather than merely asserting.

Consider the maximal, two-sided extension $r\in(-\infty,+\infty)$ of
metric~\eqref{metric} -- the Simpson--Visser spacetime itself.
Because $R(r)=R(-r)$ and $f$ depends on $r$ only through $R$, the
map $\iota:(t,r,\vartheta,\phi)\mapsto(t,-r,\vartheta,\phi)$ is an
exact isometry of the full two-sided solution, $\iota^{*}g=g$, and,
by the field equations, $\iota^{*}T^{\rm eff}=T^{\rm eff}$. We define the physical spacetime of this paper as the
$\mathbb{Z}_2$ quotient, or reflection quotient, of the maximal
extension by $\mathbb{Z}_2=\{1,\iota\}$ (since the fixed locus of
$\iota$ is a codimension-one hypersurface, this quotient is a
manifold with boundary rather than an orbifold in the strict sense).
The quotient is, tautologically, the domain $r\in[0,+\infty)$ used
throughout, and the fixed-point locus of $\iota$ is exactly the
surface $r=0$: this is the precise sense in which $r=0$ is a
boundary rather than an interior point of a larger manifold.

Crucially, this identification requires no additional matter. The
nonzero components of the extrinsic curvature of a surface
$r={\rm const}$ are proportional, up to a common finite
normalisation factor $\sqrt{|f|}$ (the modulus avoiding any formal
ambiguity in the dynamical region $f<0$, $M>M_{\min}$,
Sec.~\ref{par:dynamical_interior}), to $f_r$ and $R'$:
\begin{equation}
 K_{\vartheta\vartheta} \propto R\,R', \qquad K_{tt} \propto f_r
\end{equation}
(cf.\ the Christoffel symbols in \ref{app:Kretschmann}: $\Gamma^{r}_{\vartheta\vartheta}=-fRR'$, $f_r=2Mr/R^3$). Since $R'(r)=r/R\to0$ and $f_r=2Mr/R^3\to0$ as
$r\to0$, \emph{every} component of $K_{\mu\nu}$ vanishes at $r=0$,
for any $M$ and $\ell$ and irrespective of the sign of $f(R(0))$:
the fixed surface is \emph{totally geodesic}. By the Israel junction condition~\cite{Israel1966}, a
jump $\Delta K_{\mu\nu}$ sources a localized stress tensor; here
$\Delta K_{\mu\nu}=2K_{\mu\nu}(0)=0$ identically. The vanishing of $K_{\mu\nu}(0)$ therefore makes the $\mathbb{Z}_2$ identification \emph{tensionless}, introducing no
source beyond the effective geometric stress tensor already
computed in~\ref{app:Einstein} -- unlike Randall--Sundrum-type
orbifold fixed points, whose non-smooth warp factors
\emph{require} a brane tension~\cite{randallsundrum1999}.%
\footnote{This is a specific consequence of $R(r)$ having a smooth,
even extremum at $r=0$ (i.e.\ $R'(0)=f_r(0)=0$), not a generic
property of $\mathbb{Z}_2$ orbifold constructions: contrast the
Ho\v{r}ava--Witten $S^1/\mathbb{Z}_2$ construction of heterotic
M-theory~\cite{horavawitten1996}, whose fixed points carry $E_8$
gauge sectors and are decidedly not tensionless.}

For $M>M_{\min}$, where $r=0$ is a dynamical, spacelike bounce
rather than a static point (Sec.~\ref{par:dynamical_interior}), the
physical content of $\iota$ is a reflection: an infalling radial
trajectory that would continue smoothly through $r=0$ into the
mirror branch $r<0$ of the covering space is, in the quotient
spacetime, identified with a trajectory returning to $r>0$ --
precisely the regular bounce already identified on physical grounds
in Sec.~\ref{par:dynamical_interior}, now with an explicit geometric
origin. At no stage does the trajectory access a second
asymptotic region.

This also fixes the boundary condition for field perturbations: a
mode $\Psi$ on the covering space must satisfy
$\iota^{*}\Psi=\pm\Psi$, i.e.\ Dirichlet ($\Psi(0)=0$) or Neumann
($\Psi'(0)=0$). The energy argument of
Sec.~\ref{sec:modestability} is insensitive to this choice: the
boundary term $[\psi^{*}\dot\psi]_{r_{*}\to-\infty}$ vanishes
identically under either condition, so the mode-stability conclusion
holds in both parity sectors.
\subsection{Relation to the Bronnikov ``regular center'' program}
\label{sec:bronnikov_comparison}

Independently of, and essentially contemporaneously with, the
Simpson--Visser-descended literature discussed above, Bronnikov proposed a
different route to the same diagnosis that motivates the present
paper~\cite{bronnikov2024alternative}: that the black-bounce mechanism does
not remove the singularity at $r=0$ but merely relocates it into a wormhole
throat or a cosmological bounce, and that a genuine \emph{regular center}
should be preferred.
Restricted to the class of static, spherically symmetric metrics whose Ricci
tensor satisfies $R^{t}_{\ t}=R^{r}_{\ r}$ -- which includes Schwarzschild,
Reissner--Nordstr\"om, and Einstein--Born--Infeld -- this condition forces
the areal radius to be an affine function of the radial coordinate, so that,
without loss of generality, the areal radius coincides with the radial
coordinate itself: \emph{no deformation of the areal radius is introduced}.
Regularity is instead achieved by a Bardeen-type replacement acting only
inside the metric function; for Schwarzschild,
\begin{equation}
  A(r) = 1-\frac{2M}{r}
  \;\longmapsto\;
  A_{\rm reg}(r) = 1 - \frac{2Mr^{2}}{(r^{2}+a_{\rm Br}^{2})^{3/2}},
  \label{Bronnikov_Areg}
\end{equation}
which satisfies $A_{\rm reg}(0)=1$ and is therefore regular, by the standard
criterion adopted in~\cite{bronnikov2024alternative}, at a point $r=0$ of
\emph{vanishing} areal radius -- a center in the strict sense of that word.
The resulting metric is interpreted there as an
Einstein--nonlinear-electrodynamics solution sourced by a radial magnetic
field, with an explicit Lagrangian $\mathcal{L}(\mathcal{F})$ finite at the
center. A companion paper by Bolokhov, Bronnikov and
Skvortsova~\cite{bolokhov2024regularcenter} extends the same algorithm to the
Fisher/JNW solution and to a family of dilatonic black holes. Neither paper
addresses the thermodynamics, phase structure, or gravitational lensing of
the resulting geometries; these are left as future directions
in~\cite{bronnikov2024alternative}.

The construction of the present paper shares the interpretive goal
of~\cite{bronnikov2024alternative,bolokhov2024regularcenter} -- abandoning
the two-sided black-bounce topology in favour of a terminating boundary at
$r=0$ -- but is technically unrelated to Eq.~\eqref{Bronnikov_Areg}. Here it
is the areal radius itself, and not merely the mass function, that is
deformed, $R(r)=\sqrt{r^{2}+\ell^{2}}$ (Eq.~\eqref{Rdef}), while the
Simpson--Visser functional form $f=1-2M/R$ is left unchanged; consequently
$R(0)=\ell\neq0$, and $r=0$ is geometrically a minimal 2-sphere of finite
area $4\pi\ell^{2}$, not a point of zero area. In Bronnikov's stricter
terminology (defined above), this makes $r=0$ not a center but a regular
minimum of the areal radius -- the same local object as the
Simpson--Visser throat, now embedded in a singly connected, half-line
topology. We use ``regular center'' throughout in the topological sense
already defined in the Introduction, distinct from Bronnikov's. We go
further in Section~\ref{sec:degenerate_horizon}, where we
show explicitly that for $M>M_{\min}$ this boundary is in addition
dynamical rather than static --- the same local bounce Bronnikov
identifies within the black-hole branch $a_{\rm SV}<2m$ of the original Simpson--Visser metric~\cite{bronnikov2024alternative}; see
Section~\ref{sec:orbifold} for the precise geometric mechanism
realising this boundary as a totally geodesic $\mathbb{Z}_2$
identification. A
second structural difference concerns the source: the effective
stress-energy of the present geometry (Eqs.~\eqref{Emunu}--\eqref{EMT}) has
no independent dynamical content and no free parameter beyond $\ell$ itself,
whereas the Bronnikov construction requires an explicit
nonlinear-electrodynamics field with its own magnetic charge, which plays a
role formally analogous to $\ell$ through the NED coupling. Because of these
differences, none of the metric-, thermodynamic-, or lensing-level results
of the present paper (Sections~\ref{sec3}--\ref{sec:observational}) overlaps
with those of~\cite{bronnikov2024alternative,bolokhov2024regularcenter}; the
overlap is one of motivation and terminology.
\subsection{Related work: attribution of results on the
Simpson--Visser branch}
\label{sec:related_work}

The coincidence/difference split established in \S\ref{sec2} is
formalised, once and for all, in Table~\ref{tab:attribution}; each occurrence below (Sections~\ref{sec3}--\ref{sec:observational}) is flagged only by a bare cross-reference to
Table~\ref{tab:attribution}, without restating the comparison. An independently constructed first-law entropy also exists for the AdS embedding of the SV
metric~\cite{kumar2026simpsonvisser,noorigashti2026holographic};
both papers remain in the two-sided domain and do not identify a
degenerate-extremal endpoint, so that overlap is methodological
rather than result-level.

\paragraph{Other regular-black-hole constructions.}
Table~\ref{tab:comparison} (Section~\ref{sec:comparison}) situates
the model against other minimal-length
constructions~\cite{Nicolini2006,adler1999gravity,
maggiore1993generalized,bardeen1968non,hayward2006formation,
bronnikov2024alternative,bolokhov2024regularcenter} (the last
discussed in full in Section~\ref{sec:bronnikov_comparison}); none
share the Simpson--Visser functional form, so no metric-level
coincidence arises there.
\subsection{Horizon structure, minimal mass, and degenerate extremal
endpoint}
\label{sec:degenerate_horizon}

The horizon condition $f(R_{h})=0$ gives
\begin{equation}
  \sqrt{r_{h}^{2}+\ell^{2}} = 2M
  \quad\Longrightarrow\quad
  r_{h} = \sqrt{4M^{2}-\ell^{2}}.
  \label{rh}
\end{equation}
Three distinct mass regimes must be carefully distinguished:
\begin{itemize}
  \item $M > M_{\min} = \ell/2$: a genuine non-degenerate Killing
        horizon exists at coordinate radius $r_{h}>0$ with areal
        radius $R_{h}=2M>\ell$; the region $0\le r<r_{h}$ is
        entirely dynamical ($f<0$), not static --- see the
        paragraph below.
  \item $M = M_{\min} = \ell/2$: the coordinate horizon radius
        vanishes ($r_{h}=0$) while the areal radius remains
        $R_{h}=2M=\ell$; as established in the paragraph
        \emph{Degenerate (extremal) horizon} below, the regular
        center $r=0$ coincides \emph{exactly} with the Killing
        horizon at this mass, so the endpoint is a
        \emph{degenerate extremal regular black hole}---not a
        horizonless object.
  \item $M < M_{\min}$: no horizon forms; the geometry describes a
        regular horizonless compact object, static throughout
        ($f>0$ everywhere), with a genuine Bardeen/Hayward-type
        static regular center at $r=0$ --- see the paragraph below.
\end{itemize}

\paragraph{Nature of the region $0\le r<r_{h}$ for $M>M_{\min}$: a
dynamical bounce, not a static core.}
\label{par:dynamical_interior}

Differentiating $f(R(r))=1-2M/R(r)$ directly with respect to $r$
recovers the quantity $f_r=2Mr/R^{3}$ already introduced in~\ref{app:Kretschmann}:
\begin{equation}
  \frac{d}{dr}\bigl[f(R(r))\bigr]
  = \frac{2M}{R^{2}}\frac{dR}{dr}
  = \frac{2Mr}{R^{3}} \;\geq\; 0 ,
  \label{fmonotone}
\end{equation}
with equality only at $r=0$: $f\circ R$ is therefore
\emph{strictly monotonically increasing} on $[0,+\infty)$, running
from $f(R(0))=1-2M/\ell$ to $f(\infty)=1$. This monotonicity has three immediate consequences.

\begin{itemize}
  \item \emph{Exactly one horizon.} Since $f\circ R$ is strictly
    monotonic, it can vanish at most once on $[0,+\infty)$: the
    single root $r_{h}=\sqrt{4M^{2}-\ell^{2}}$ of Eq.~\eqref{rh} is
    therefore the \emph{only} horizon of the geometry, for any $M$.
    Unlike a Bardeen- or Hayward-type regular black hole, $f$ never
    turns positive again between an inner and an outer horizon.
  \item \emph{Sign of $f$ at the boundary.} For $M>M_{\min}=\ell/2$,
    $f(R(0))=1-2M/\ell<0$; for $M<M_{\min}$, $f(R(0))>0$; for
    $M=M_{\min}$, $f(R(0))=0$ exactly, which re-derives
    Eq.~\eqref{f_at_center}.
  \item \emph{Character of the region $0\le r<r_{h}$.} Whenever
    $M>M_{\min}$, monotonicity forces $f(R(r))<0$ throughout the
    \emph{entire} interval $0\le r<r_{h}$, with no sub-interval in
    which $f$ returns to positive values.
\end{itemize}

This point requires a physical reading. Since
$g_{tt}=f(R)$ and $g_{rr}=-f(R)^{-1}$, the region $f<0$ is one in
which $t$ is spacelike and $r$ is timelike: exactly as in the
interior of classical Schwarzschild, the hypersurfaces
$r=\mathrm{const}$ are spacelike there, and the induced metric on
them is that of a spatially homogeneous, anisotropic
Kantowski--Sachs-type cosmology, in which the areal radius $R(r)$
plays the role of a time-dependent scale factor. Since $dR/dr=r/R
\to 0$ while $R(r)$ reaches a finite, smooth minimum
$R_{\min}=\ell$ at $r=0$ (Eq.~\eqref{f_quadratic} below makes the
same observation for $f$ itself), the point $r=0$ is reached
\emph{dynamically}, as the turning point (bounce) of a
contracting-then-expanding scale factor, not as a static core
reached while $f>0$ -- the same conclusion Bronnikov reaches for
the black-hole branch $a_{\rm SV}<2m$ of the original
Simpson--Visser metric~\cite{bronnikov2024alternative}
(Section~\ref{sec:bronnikov_comparison}).

The qualifier ``regular center'' should therefore \emph{not} be
read as implying a static Bardeen/Hayward-type core for
$M>M_{\min}$: in that regime, $r=0$ is more accurately described as
a \emph{regular bounce} of the areal radius inside a dynamical
interior, exactly analogous to the interior bounce of
Simpson--Visser itself. A genuine static core in the
Bardeen/Hayward sense is realised in this geometry only for
$M<M_{\min}$, where $f(R(r))>0$ everywhere; at $M=M_{\min}$ the
bounce and the horizon coincide exactly
(Section~\ref{sec:degenerate_horizon}). Section~\ref{sec:orbifold}
gives the precise geometric mechanism underlying both regimes: the
totally geodesic $\mathbb{Z}_2$ identification at $r=0$ is
approached through $f<0$ (dynamical bounce) or $f>0$ (static core)
according to the sign of $f(R(0))$, so the two behaviours are the
same fixed-point structure read in two different mass ranges, not
two different constructions.
\paragraph{Degenerate (extremal) horizon at $M=M_{\min}$.}

At $M = M_{\min}=\ell/2$:
$R(0)=\ell = 2M = R_{h}$, so the regular center $r=0$ lies
\emph{exactly on} the horizon locus.
One verifies directly that
\begin{equation}
  f\!\bigl(R(r=0)\bigr)\big|_{M=\ell/2}
  = 1 - \frac{2(\ell/2)}{\ell} = 0,
  \label{f_at_center}
\end{equation}
confirming that $r=0$ is a \emph{Killing horizon}: the norm of
$\partial_{t}$ vanishes there, and $g^{rr}=-f(R)=0$, so the normal
to the surface $r=\mathrm{const}$ is null.
The remnant at $M=M_{\min}$ is therefore \emph{not} a horizonless
object; it possesses a genuine (degenerate) horizon.

The nature of this horizon is revealed by expanding $f$ near $r=0$
at $M=\ell/2$:
\begin{equation}
  R(r)\big|_{r\to 0} = \sqrt{r^{2}+\ell^{2}}
  \approx \ell\Bigl(1+\frac{r^{2}}{2\ell^{2}}\Bigr),
\end{equation}
so that
\begin{equation}
  f(R)\big|_{M=\ell/2}
  = 1 - \frac{\ell}{R}
  \approx 1 - \frac{1}{1+r^{2}/(2\ell^{2})}
  \approx \frac{r^{2}}{2\ell^{2}}
  \quad (r\to 0).
  \label{f_quadratic}
\end{equation}
This \emph{quadratic vanishing} $f\sim (r-r_{h})^{2}$ with
$r_{h}=0$ is characteristic of a \emph{degenerate} (extremal) horizon---precisely the same structure as the coincident inner
and outer horizons of an extremal Reissner--Nordstr\"{o}m black
hole~\cite{Wald1994,chen2015black}.
The surface gravity $\kappa = r_{h}/(8M^{2})\to 0$ as $r_{h}\to 0$,
yielding $T_{H}\to 0$, fully consistent with the extremal interpretation.

We therefore characterise the evaporation endpoint as a
\emph{degenerate extremal regular black hole}: a configuration in
which the regular center $r=0$ and the Killing horizon coincide at
the areal radius $\ell$, with vanishing surface gravity, vanishing
temperature, and finite curvature everywhere.
The terminology ``regular remnant'' or ``extremal remnant'' will
be used interchangeably throughout; the qualifier ``horizonless''
employed in some earlier analyses of related metrics is
\emph{incorrect} at $M=M_{\min}$ and is avoided in the present paper.

\subsection{Global causal structure}
\label{sec:penrose}

\begin{figure}[!htbp]
\centering
\includegraphics[width=\linewidth]{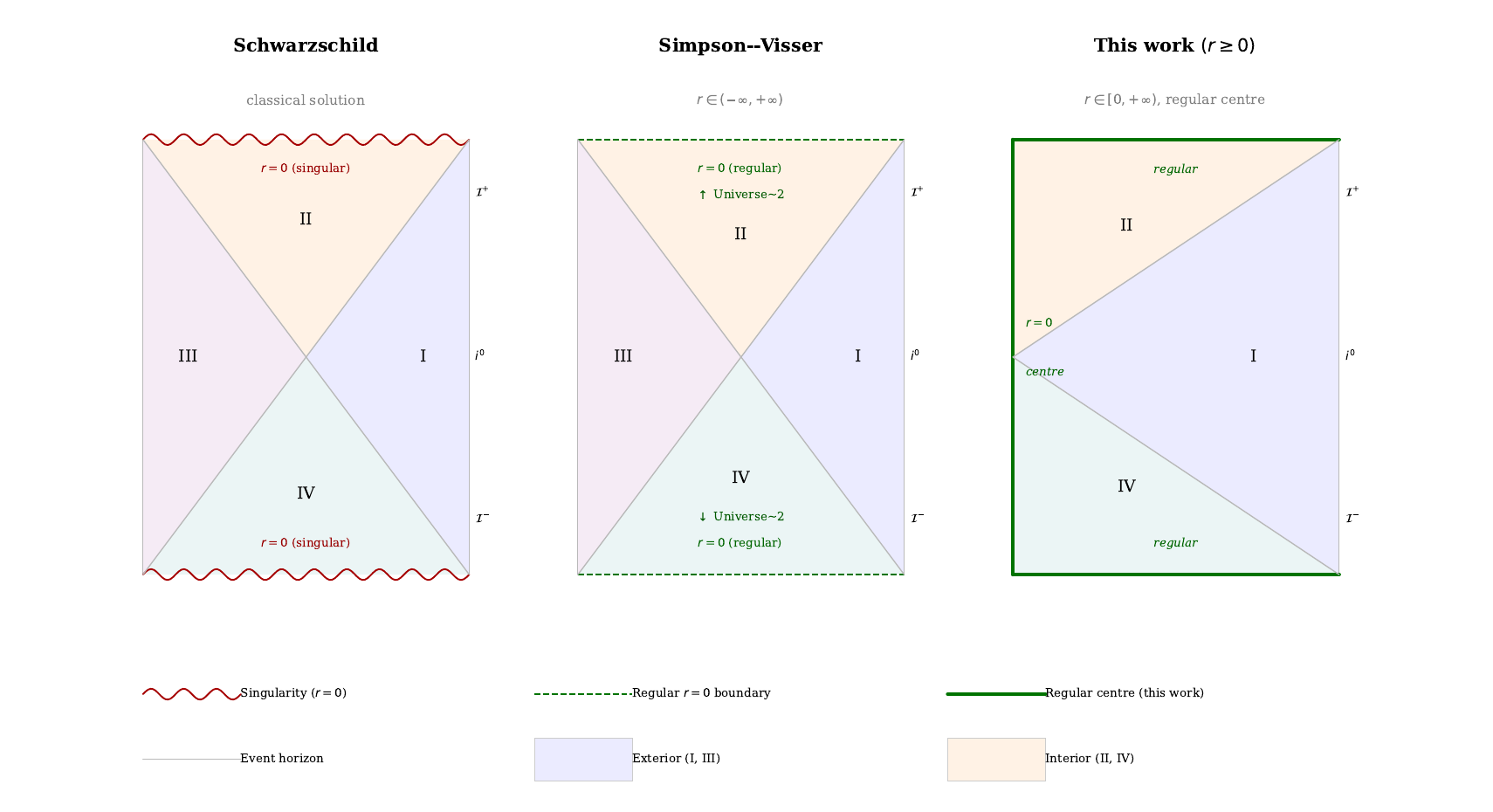}
\caption{Qualitative Penrose--Carter diagrams for the three
  spacetimes (see Sec.~\ref{sec:penrose} for the discussion).
  Left: Schwarzschild, with spacelike singularities and a second
  exterior Region~III. Centre: Simpson--Visser ($r\in\mathbb{R}$),
  where the regular $r=0$ surface is a wormhole throat. Right:
  this work ($r\ge0$), where $r=0$ is a terminating regular
  centre and Region~III is absent. As $M\to M_{\min}$, Regions~II
  and~IV shrink and the horizon degenerates
  (Sec.~\ref{sec:degenerate_horizon}). In the right-hand panel, the
  vertical segment is a schematic device marking the locus of the
  totally geodesic $\mathbb{Z}_2$ fixed point (Sec.~\ref{sec:orbifold})
  that replaces Region~III's boundary; it does not indicate the
  causal character of $r=0$, which remains spacelike throughout
  $0\le r<r_{h}$ for $M>M_{\min}$ (Sec.~\ref{sec:degenerate_horizon}).}
\label{fig:penrose}
\end{figure}

The restriction $r\geq 0$ has a direct consequence for the maximal analytic extension of the metric.
Figure~\ref{fig:penrose} displays qualitative Penrose--Carter conformal
diagrams for the three spacetimes under comparison, making explicit
the causal differences that result from different treatments of the
$r=0$ surface.

\paragraph{Classical Schwarzschild (Fig.~\ref{fig:penrose}, left).}
The maximal Kruskal--Szekeres extension contains four regions: the
right exterior Region~I, the future black-hole interior Region~II
bounded by the spacelike singularity $r=0$ (wavy red lines), the
left mirror exterior Region~III, and the past white-hole interior
Region~IV.
Both singularities at the top and bottom of the diagram constitute
the inextendible causal boundary of all future- and
past-directed geodesics that enter the interior.

\paragraph{Simpson--Visser (Fig.~\ref{fig:penrose}, centre).}
Allowing $r\in(-\infty,+\infty)$ replaces both singularities by
\emph{regular} $r=0$ surfaces (dashed green lines).
The topology becomes that of a black-bounce wormhole: Regions~I
and~III are two distinct asymptotically flat universes connected
through the regular $r=0$ surface acting as a minimal throat.
The arrows labelled ``Universe~2'' in Fig.~\ref{fig:penrose} indicate
that interior Regions~II and~IV each communicate through this throat
with a mirror universe~\cite{simpson2019black,lobo2021ultracompact}.
The global topology is doubly connected.

\paragraph{This work (Fig.~\ref{fig:penrose}, right).}
Restricting $r\geq 0$ eliminates Region~III entirely: the regular
center introduced in \S\ref{sec2} now plays the role of a smooth,
simply connected boundary, in place of the singularity of
Schwarzschild or the wormhole throat of Simpson--Visser.
For $M>M_{\min}$, the center lies strictly inside the horizon
($R(0)=\ell < 2M = R_{h}$), where it is reached dynamically as a
spacelike bounce rather than as a static core
(Sec.~\ref{par:dynamical_interior} gives the argument).
The maximal extension contains only:
\begin{itemize}
  \item Region~I: the single asymptotically flat exterior,
  \item Region~II: the future black-hole interior, bounded
        \emph{above} by the regular center (thick green line),
  \item Region~IV: the time-reversed past white-hole interior,
        bounded \emph{below} by the same regular center.
\end{itemize}
The thick solid green line in Fig.~\ref{fig:penrose} (right) marks
this regular boundary, in contrast to the singular
boundary of Schwarzschild and to the connecting wormhole throat
of Simpson--Visser.

As Hawking evaporation drives $M\to M_{\min}=\ell/2$
(Section~\ref{sec3}), Region~II progressively shrinks.
At the endpoint $M=M_{\min}$, the horizon degenerates: the regular
center $r=0$ merges with the event horizon at the areal radius $\ell$
(Eqs.~\eqref{f_at_center}--\eqref{f_quadratic}), and the interior
Regions~II and~IV shrink to zero, leaving a
\emph{degenerate extremal regular black hole}.
This limiting configuration is not horizonless---it is an
extremal-type object whose Penrose diagram is the degenerate limit
in which the two diagonal lines coincide at the extremal bifurcation
surface $r=0$, analogous to the degenerate diagram of an extremal
Reissner--Nordstr\"{o}m black hole.

Three key topological features distinguish the present spacetime
from its competitors and are directly visible in
Fig.~\ref{fig:penrose}:
\begin{enumerate}
  \item \textit{No singularity.}
        The wavy red lines of Schwarzschild are absent;
        $r=0$ is a regular hypersurface with finite curvature
        invariants (Section~\ref{sec:curv}).
  \item \textit{No second exterior.}
        Unlike both Schwarzschild and Simpson--Visser, there is
        no Region~III; the spacetime has a single asymptotic
        end and is simply connected.
  \item \textit{No wormhole throat.}
        Unlike Simpson--Visser, the regular
center is a terminating boundary, not a passage to another
universe (Sec.~\ref{sec:degenerate_horizon}).
\end{enumerate}

\subsection{Curvature regularity}
\label{sec:curv}

Computing the Kretschmann scalar (see~\ref{app:Kretschmann}
for the complete derivation) yields
\begin{equation}
  K(r) = \frac{4\bigl[
    M^{2}(12R^{4}-36\ell^{2}R^{2}+33\ell^{4})
    +8M\ell^{2}R(R^{2}-2\ell^{2})
    +3\ell^{4}R^{2}
  \bigr]}{R^{10}},
  \label{K_exact}
\end{equation}
where $R = \sqrt{r^{2}+\ell^{2}}$.
This expression is finite everywhere on $r\in[0,+\infty)$, with
\begin{equation}
  K(0) = \frac{4(9M^{2}-8M\ell+3\ell^{2})}{\ell^{6}},
  \label{K0}
\end{equation}
confirming the removal of the classical curvature singularity.
Regularity holds even at the degenerate extremal endpoint
$M=\ell/2$, where $K(0)=4(9/4-4+3)/\ell^{4} = 4(5/4)/\ell^{4}
=5/\ell^{4}$, which is finite.
In the asymptotic regime $r\gg\ell$ one recovers
$K \to 48M^{2}/r^{6}$, the standard Schwarzschild result.

\subsection{Effective source and its geometric origin}

For $\ell>0$, the Einstein tensor of metric~\eqref{metric}
is nonvanishing (\ref{app:Einstein}):
\begin{equation}
  G^{t}_{\ t} = \frac{\ell^{2}(4M-R)}{R^{5}},\quad
  G^{r}_{\ r} = \frac{\ell^{2}}{R^{4}},\quad
  G^{\vartheta}_{\ \vartheta}=G^{\phi}_{\ \phi}
    = \frac{\ell^{2}(M-R)}{R^{5}}.
  \label{Emunu}
\end{equation}
All components vanish as $\ell\to 0$, confirming that the source
is of purely geometric origin with no classical matter content.
The effective energy-momentum tensor
$T^{\mu}_{\ \nu}=G^{\mu}_{\ \nu}/(8\pi)$
describes an anisotropic fluid with

\begin{equation}
  \rho = \frac{\ell^{2}(4M-R)}{8\pi R^{5}}, \quad
  p_{r} = -\frac{\ell^{2}}{8\pi R^{4}}, \quad
  p_{\vartheta} = \frac{\ell^{2}(R-M)}{8\pi R^{5}}.
  \label{EMT}
\end{equation}
The effective source satisfies the Ricci scalar
\begin{equation}
  \mathcal{R}_{\text{sc}} = \frac{2\ell^{2}(R-3M)}{R^{5}},
  \label{Ricciscalar}
\end{equation}
which vanishes at $R=3M$ (photon-sphere radius, established in
Section~\ref{sec:shadow}) and in the limit $\ell\to 0$.

\paragraph{Remark on the label ``purely geometric source.''}
We use this term in a specific and limited sense that deserves
explicit clarification.
Mathematically, every metric satisfies
$G_{\mu\nu}=8\pi T^{\rm eff}_{\mu\nu}$ by the contracted Bianchi
identity; calling the right-hand side ``geometric'' rather than
``exotic matter'' is an interpretational choice, not a mathematical
property of the field equations.
What justifies the geometric label \emph{in the present context}
is the following set of properties:
(a)~$T^{\rm eff}_{\mu\nu}$ has no independent dynamical degree of
freedom---it is algebraically determined by the metric and the
single deformation parameter $\ell$, with no propagating matter
field and no separate equation of motion;
(b)~it vanishes \emph{identically} when $\ell\to 0$
(Eq.~\eqref{Emunu} above), recovering the Schwarzschild vacuum
without any residual matter contribution;
(c)~it arises solely from the minimal deformation of the
areal-radius coordinate $r\mapsto R(r)=\sqrt{r^2+\ell^2}$,
encoding a Planck-scale modification of the geometry itself
rather than coupling to a matter Lagrangian.
In this sense, the source is ``geometric'' in the same spirit that
the Ricci scalar in $f(R)$ gravity generates an effective
stress-energy without coupling to a separate matter field.
Nevertheless, the effective fluid does violate standard energy
conditions in part (Section~\ref{sec:energy}), a property shared by all
known regular black hole geometries~\cite{barcelo2000scalar,
hayward2006formation,Nicolini2006}, and is therefore ``exotic''
in the energy-condition sense regardless of its geometric origin.
Both descriptions---geometric deformation and effective exotic
fluid---are mathematically equivalent via
$G_{\mu\nu}=8\pi T^{\rm eff}_{\mu\nu}$; we use the geometric
language throughout as a reminder that $T^{\rm eff}_{\mu\nu}$
carries no independent dynamics and vanishes in the classical limit.

Energy conditions are analysed in Section~\ref{sec:energy}.

\section{Horizons and Thermodynamics}
\label{sec3}

\subsection{Hawking temperature}
\label{sec:Temp}

Since $\sqrt{-g_{tt}g_{rr}}=1$ identically for
metric~\eqref{metric} (so the zeroth law holds automatically), the
surface gravity reduces directly to $\kappa=Mr_h/R_h^3=r_h/(8M^2)$,
giving the Hawking temperature
\begin{equation}
  T_{H} = \frac{\kappa}{2\pi} = \frac{r_{h}}{16\pi M^{2}}
    = \frac{\sqrt{4M^{2}-\ell^{2}}}{16\pi M^{2}}.
  \label{TH}
\end{equation}
This is the known Simpson--Visser Hawking
temperature~\cite{simpson2019black,joshi2026thermodynamic}
(cf.\ Table~\ref{tab:attribution}). We retain only the two values
needed below: the maximum
\begin{equation}
  T_{H}^{\max} = T_{H}(M^{*}) = \frac{1}{8\pi\ell},
  \qquad M^{*} = \ell/\sqrt{2},
  \label{THmax}
\end{equation}
and the vanishing $T_{H}\to 0$ as $M\to M_{\min}^{+}=(\ell/2)^{+}$,
which fixes the lower boundary of the entropy integration in
\S\ref{sec:BC}.

\subsection{Entropy from the first law}

We derive the entropy directly from the first law of black hole
thermodynamics $dM = T_{H}\,dS$, using $M$ as the thermodynamic
energy.
This identification is self-consistent: the ADM mass $M$ is the
conserved charge associated with time-translation invariance and is
the appropriate thermodynamic energy for the canonical ensemble.

From the horizon relation~\eqref{rh}, differentiating with respect
to $r_{h}$:
\begin{equation}
  \frac{dM}{dr_{h}} = \frac{r_{h}}{2R_{h}}, \qquad
  R_{h} = \sqrt{r_{h}^{2}+\ell^{2}}.
  \label{dMdrh}
\end{equation}
From Eq.~\eqref{TH}, the temperature can be written as
\begin{equation}
  T_{H} = \frac{r_{h}}{4\pi R_{h}^{2}},
\end{equation}
which follows since $16\pi M^{2} = 4\pi R_{h}^{2}$ at the horizon.
The first law then gives
\begin{equation}
  \frac{dS}{dr_{h}} = \frac{1}{T_{H}}\frac{dM}{dr_{h}}
  = \frac{4\pi R_{h}^{2}}{r_{h}} \cdot \frac{r_{h}}{2R_{h}}
  = 2\pi R_{h} = 2\pi\sqrt{r_{h}^{2}+\ell^{2}}.
  \label{dSdrh}
\end{equation}
This result has a simple closed form: the entropy gradient with
respect to the horizon coordinate equals $2\pi$ times the areal
radius, the natural geometric quantity associated with the deformed
2-sphere.
This result does \emph{not} reduce to the Bekenstein--Hawking area
law when re-expressed in terms of the areal radius $R_h$ itself.
Using $dR_h/dr_h=r_h/R_h$, Eq.~\eqref{dSdrh} is equivalent
to
\begin{equation}
  \frac{dS}{dR_{h}} = \frac{dS/dr_h}{dR_h/dr_h}
  = \frac{2\pi R_h^{2}}{r_h},
  \label{dSdRh}
\end{equation}
which differs from the area-law derivative $dS_{\rm area}/dR_h=2\pi
R_h$ for any $\ell>0$ (the two coincide only as $\ell\to0$, where
$r_h\to R_h$); indeed Eq.~\eqref{dSdRh} diverges as $r_h\to0$,
whereas $2\pi R_h$ stays finite— a
direct consequence of the nature of the effective source. For pure
Einstein--Hilbert gravity, Wald's Noether-charge construction
guarantees that the geometric entropy associated with \emph{any}
Killing horizon is exactly $S_{\rm Wald}=A/4$, independently of the
matter or effective source present~\cite{wald1993black,
iyer1994some}; the identification of this quantity with the
thermodynamic entropy appearing in the first law -- whether derived
in the equilibrium-state sense of Iyer and Wald~\cite{iyer1994some}
or in the physical-process sense of Bardeen, Carter and
Hawking~\cite{bardeen1973four} -- relies in either case on
properties of the matter/effective-source content (satisfaction of
the relevant energy conditions, or, equivalently, the existence of a
well-posed matter action whose parameter dependence supplies the
correct work term) that are not guaranteed to hold for an arbitrary
effective source. As shown explicitly in Section~\ref{sec:energy}
(Eq.~\eqref{NECr}), the effective stress-energy sourcing the present
geometry violates the radial null energy condition throughout the
exterior $R>2M$; this is precisely the type of violation that is
known, in the regular-black-hole literature, to decouple the entropy
consistent with $dM=T_H\,dS$ from the bare area law, replacing
$S=A/4$ by a source-dependent, generally smaller
entropy~\cite{maZhao2014correctedfirstlaw}. Equation~\eqref{Entropy}
should therefore be understood as the entropy consistent with the
equilibrium relation $dM=T_H\,dS$ at fixed $\ell$ -- not as the
Wald/area-law entropy of the same horizon, which remains $\pi R_h^2$
throughout and does not vanish at the degenerate endpoint (see
Section~\ref{sec:BC} below) -- and the discrepancy between the two
is expected precisely because of the NEC violation just described,
rather than being an artefact of the calculation.
Equation~\eqref{dSdrh} is a purely metric statement
(Section~\ref{sec:related_work}); integrating it under the boundary condition justified in
\S\ref{sec:BC} reproduces the corresponding term of Joshi and
Joshi's entropy (\S\ref{sec:entropy_SV_comparison}).
Integrating from $r_{h}=0$ (where $S=0$, justified below)
to $r_{h}$:
\begin{align}
  S(r_{h}) &= \int_{0}^{r_{h}} 2\pi\sqrt{r^{\prime 2}+\ell^{2}}\,dr^{\prime}
  \nonumber\\
  &= \boxed{\pi\left[r_{h}\sqrt{r_{h}^{2}+\ell^{2}}
      + \ell^{2}\ln\!\left(\frac{r_{h}+\sqrt{r_{h}^{2}+\ell^{2}}}{\ell}
        \right)
    \right]}.
  \label{Entropy}
\end{align}
This is the main result for the entropy of the minimally deformed
Schwarzschild black hole, subject to the boundary condition motivated and discussed
critically in Section~\ref{sec:BC}.

\paragraph{A note on novelty, stated here rather than deferred.}
The functional form of Eq.~\eqref{Entropy} coincides, under
$\ell\leftrightarrow a_{\rm SV}$, with the semiclassical entropy
term identified independently by Joshi and
Joshi~\cite{joshi2026thermodynamic}, once their integration constant
is fixed by the same extremal boundary condition we adopt in
\S\ref{sec:BC}; this is expected, since both derivations integrate
the same first law $dS=dM/T_H$ from the same shared $T_H(M,\ell)$
(\S\ref{sec:entropy_SV_comparison} gives the full argument). What is
new here is not this functional form but (i)~its independent,
purely geometric boundary-condition justification in \S\ref{sec:BC},
tied to the degenerate extremal horizon of
\S\ref{sec:degenerate_horizon} rather than to a wormhole--black-hole
transition; (ii)~its adoption as the \emph{complete} physical
entropy rather than a semiclassical piece awaiting tunneling
corrections; and (iii)~the free energy $F(M)$
(\S\ref{sec:freeenergy}), the geometrothermodynamic analysis
(\S\ref{sec:GTD}), and the mode-stability result
(\S\ref{sec:modestability}) built on it, none of which appear
in~\cite{joshi2026thermodynamic}. See \S\ref{sec:entropy_SV_comparison}
for the itemised comparison.

\subsubsection{Boundary condition \texorpdfstring{$S(r_{h}=0)=0$}{S(rh=0)=0} and its physical
justification.}
\label{sec:BC}

At $r_{h}=0$, the coordinate horizon radius vanishes, but a
\emph{degenerate} Killing horizon persists at $r=0$ with areal
radius $\ell$ (\S\ref{sec:degenerate_horizon}).
Formula~\eqref{Entropy} satisfies $S(0)=0$ \emph{automatically}:
$r_{h}R_{h}\to 0$ and $\ell^{2}\ln(\ell/\ell)=0$.

This zero entropy at the degenerate extremal endpoint is supported
by two physical considerations (\S\ref{sec:degenerate_horizon}):

\begin{enumerate}
  \item \textit{Third law and microstate degeneracy.}
  The analogue of Nernst's postulate for black holes states that
  any process which drives the surface gravity (and hence the
  temperature) to zero must also drive the entropy to
  zero~\cite{Wald1994}: in our model $T_{H}\to 0$ as
  $M\to M_{\min}^{+}=(\ell/2)^{+}$ because $\kappa=r_{h}/(8M^{2})\to0$,
  so the third law requires $S\to0$ at the endpoint; had the
  integration in Eq.~\eqref{Entropy} started at any $r_{h}>0$, this
  would fail, which fixes the lower bound. The same conclusion
  follows independently from microstate counting: in the
  Bekenstein--Hawking framework~\cite{Bekenstein}, the degenerate
  extremal endpoint ($T_{H}=0$) is uniquely specified by $\ell$
  alone, with no thermally accessible microstate excitations, so by
  analogy with extremal Reissner--Nordstr\"{o}m black holes the
  ground-state degeneracy is one and $S=\ln1=0$~\cite{chen2015black}.
  These are two readings -- thermodynamic and statistical -- of the
  same expectation that a horizon of vanishing surface gravity
  carries vanishing entropy, rather than logically independent
  derivations.

  Hawking, Horowitz and Ross~\cite{hawkinghorowitzross1995} showed
  that for the extremal Reissner--Nordstr\"om black hole, continuity
  of the area-law entropy across the near-extremal family gives
  instead a \emph{nonzero} limit $S\to\pi Q^2$ as $T_H\to0$; that
  case is not, however, on the same footing as ours, since the
  Maxwell source satisfies the NEC and $S=A/4$ holds exactly for
  every $r_h>0$ there, whereas here the radial NEC is violated
  throughout the exterior (Section~\ref{sec:energy},
  Eq.~\eqref{NECr}), so -- as explained above, Eq.~\eqref{dSdRh} --
  there is no expectation that the first-law entropy of
  Eq.~\eqref{Entropy} should agree with the area-law value
  $\pi R_h^2$ in the first place; indeed $\pi R_h^2\to\pi\ell^2\neq0$
  at the degenerate endpoint even though $S\to0$ in
  Eq.~\eqref{Entropy}. We adopt the first-law-consistent entropy of
  Eq.~\eqref{Entropy}, rather than the area-law value, as the
  physical entropy of the model; the
  Hawking--Horowitz--Ross-type ambiguity would resurface only if the
  area-law value were adopted instead, a choice we explicitly do not
  make.

  \item \textit{Degenerate-extremal horizon and the Euclidean
  path integral.}
  For an extremal horizon, the inverse temperature
  $\beta = T_{H}^{-1}\to\infty$: Euclidean time becomes
  non-compact and the on-shell gravitational action vanishes
  identically~\cite{Wald1994,hawkingpage1983}, so the gravitational
  partition function reduces to a trivial saddle with zero action,
  giving $S = -\partial F/\partial T \to 0$ consistently with the
  first-law derivation -- the standard result for degenerate
  extremal black holes (see e.g.\ Refs.~\cite{chen2015black,Wald1994}).
  This is the generic extremal-horizon result, applied here by
  analogy rather than confirmed from the model-specific on-shell
  action $I_E(r_h,\ell)$; an explicit computation showing
  $I_E\to0$ continuously as $r_h\to0$ is left for future work
  (Sec.~\ref{sec:future}).
\end{enumerate}

Together with the argument above for why the area-law value is not
expected to apply here in the first place, these two considerations
support -- but do not uniquely force -- the boundary condition
$S(r_h=0)=0$ on which Eq.~\eqref{Entropy} and the free energy, GTD,
and stability analyses of Sections~\ref{sec4} and~\ref{sec:GTD} are
built.
\paragraph{Scope of the claim relative to the on-shell calculation
of Sec.~\ref{sec:future}.}
Section~\ref{sec:future} lists a first-principles computation of
the on-shell Euclidean action $I_E(r_h,\ell)$ for this metric as an
open problem, and it is worth clarifying explicitly why
Eq.~\eqref{Entropy}, so normalised, is already adopted here as the
complete physical entropy of the model rather than held provisional
pending that calculation.

Both arguments given above are generic statements about \emph{any}
horizon of vanishing surface gravity, not statements requiring
prior knowledge of $I_E(r_h,\ell)$ for the present effective source
specifically: the third-law argument depends only on $\kappa\to0$,
already established in closed form in Sec.~\ref{sec:Temp}, and the
Euclidean argument depends only on the fact that
$\beta=T_H^{-1}\to\infty$ trivialises the on-shell action of any
smooth degenerate saddle. The calculation flagged in
Sec.~\ref{sec:future} would make the approach to $S=0$ quantitative
and model-specific, but there is no way for it to remain consistent
with $\kappa\to0$ and yet select $S\neq0$; it would refine, not
overturn, the conclusion reached here.

Fixing a first-law integration constant by third-law reasoning,
ahead of an explicit on-shell derivation, is in any case the
standard order of operations for extremal and near-extremal objects
in this literature. Indeed, even where an explicit calculation
\emph{is} available, as for extremal Reissner--Nordstr\"{o}m, its
physical reading is not settled by the calculation alone: Hawking,
Horowitz and Ross~\cite{hawkinghorowitzross1995} obtain a nonzero
limit $S\to\pi Q^2$ from area-law continuity, in tension with the
third-law expectation invoked above (\S\ref{sec:BC}). An eventual
computation of $I_E(r_h,\ell)$ for the present model would
therefore still need to be interpreted against exactly this kind of
physical argument, rather than replacing it.

Finally, ``complete physical entropy'' is not intended here as a
claim of first-principles proof. As already stated in the
Introduction, the boundary condition is motivated on physical
grounds rather than independently proven; Eq.~\eqref{Entropy} is
the entropy \emph{consistent with this physically motivated
choice}, and it is on this basis -- not on an as-yet-unperformed
on-shell calculation -- that the free energy, GTD, and stability
analyses of Sections~\ref{sec4}--\ref{sec:GTD} are built. Should the
calculation of Sec.~\ref{sec:future} eventually be carried out, we
expect it to confirm this picture quantitatively rather than to
alter it.
\subsubsection{Relation to the entropy of Joshi and Joshi, and to
Simpson--Visser entropy analyses on \texorpdfstring{$r\geq 0$}{r >= 0}.}
\label{sec:entropy_SV_comparison}

\paragraph{The semiclassical entropy formula is shared, and
necessarily so.}
As already noted in Section~\ref{sec3}, substituting
$r_h=\sqrt{4M^2-\ell^2}$ and $R_h=2M$ into Eq.~\eqref{Entropy} gives
\begin{equation}
  S(M) = 2\pi\left[M\sqrt{4M^2-\ell^2}
    + \frac{\ell^2}{2}\,
      \ln\!\left(\frac{2M+\sqrt{4M^2-\ell^2}}{\ell}\right)\right].
  \label{S_of_M}
\end{equation}
Under $\ell\leftrightarrow a_{\rm SV}$, this is algebraically
identical to the term Joshi and Joshi denote $S_0$ -- the leading
(semiclassical) piece of their quantum-corrected
entropy~\cite[Eq.~(24)]{joshi2026thermodynamic} -- once their
integration constant is fixed by the same requirement they impose,
$S\to0$ as $M\to\ell/2$~\cite[Eq.~(28)]{joshi2026thermodynamic}; the
match is with this boundary-condition-fixed form, combined into
their final Eq.~(27), not with the unnormalised Eq.~(24) in
isolation. This is not a coincidence: $T_H(M,\ell)$ is the same
function in both papers (Sec.~\ref{sec:related_work}), the first law
$dS=dM/T_H$ is a first-order ODE, and a fixed boundary value
determines its solution uniquely, so any derivation sharing $T_H$
and the same extremal boundary condition must return the same
$S(M)$. Joshi and Joshi make the same point about this shared term:
thermodynamic integration of the first law already introduces an
$\ell$-dependence at semiclassical order, in contrast to the
$\ell$-independent area-law value $S_{BH}=4\pi M^2$ (their $m\equiv$
our $M$). What distinguishes the present analysis is therefore
narrower than the functional form itself:
\begin{enumerate}
 \item \emph{Status of the entropy, and physical origin of the
  shared boundary condition.} Joshi and Joshi treat
  Eq.~\eqref{S_of_M} as the leading piece of a further
  quantum-corrected entropy, obtained by superposing
  Hamilton--Jacobi tunneling corrections
  $\propto\beta_1,\beta_2$~\cite[Eqs.~(25)--(27)]{joshi2026thermodynamic}.
  In their two-sided construction, $S\to0$ reflects $r_h=0$ marking
  the SV \emph{wormhole--black-hole transition}: for $M<\ell/2$ the
  metric describes a traversable wormhole, so $r_h\to0^+$ means
  approaching the wormhole regime rather than a Killing horizon. We
  do not include tunneling corrections: on the strength of the
  three arguments of \S\ref{sec:BC}, which instead identify $r_h=0$
  as a genuine, degenerate Killing horizon of a \emph{simply
  connected} spacetime (Section~\ref{sec:degenerate_horizon}), we
  adopt Eq.~\eqref{Entropy} as the complete physical entropy of the
  degenerate extremal remnant -- giving the shared boundary
  condition an independent physical origin, without altering
  $S(M)$ itself.
  \item \emph{Which entropy enters the free energy.} The canonical
  free energy of~\cite[Eq.~(15)]{joshi2026thermodynamic} is built on
  the bare area law $S_{BH}=4\pi M^2$, not on Eq.~\eqref{S_of_M}; our
  free energy $F(M)$ of Eq.~\eqref{HelmholtzF} is instead built on
  Eq.~\eqref{Entropy}, so the resulting phase selection of
  \S\ref{sec:freeenergy}, together with the geometrothermodynamic
  and mode-stability analyses of Sections~\ref{sec:GTD}
  and~\ref{sec:modestability}, remain independent contributions.
\end{enumerate}
Table~\ref{tab:attribution} reflects this narrower statement.

\paragraph{Physical reinterpretation of the linear entropy correction.}
The small-horizon series~\eqref{S-smallrh},
$S\approx 2\pi\ell\,r_{h}+\pi r_{h}^{3}/(3\ell)+\ldots$, follows from
the even symmetry of $R_{h}(r_{h})$ about $r_{h}=0$ and from
$S(0)=0$. We read $S\approx 2\pi\ell\,r_{h}$ as the
\emph{near-extremal entropy} of the degenerate configuration -- the
horizon is a thin annular shell of areal radius $R_{h}\approx\ell$
-- recovering the quadratic area law only for $r_{h}\gg\ell$, and
tie it explicitly to the degenerate extremal remnant of
Section~\ref{sec:degenerate_horizon}. This reading does not appear
in~\cite{joshi2026thermodynamic}.
\subsubsection{Classical limit.}

As $\ell\to 0$: $r_{h}R_{h}\to r_{h}^{2}$ and
$\ell^{2}\ln(\cdots)\to 0$, giving
$S\to\pi r_{h}^{2} = A/4$, the Bekenstein--Hawking area law.

\subsubsection{Corrections for small horizons
\texorpdfstring{($r_{h}\ll\ell$)}{(r\_h << ell)}.}
\label{sec:smallrh}

The Taylor expansion of $S(r_h)$ about $r_h=0$ follows directly from
the entropy gradient $S'(r_{h}) = 2\pi\sqrt{r_{h}^{2}+\ell^{2}}$
and its successive derivatives evaluated at $r_h=0$:
\begin{equation}
  S'(0) = 2\pi\ell, \qquad
  S''(0) = \frac{2\pi r_{h}}{\sqrt{r_{h}^{2}+\ell^{2}}}\bigg|_{0} = 0,
  \qquad
  S'''(0) = \frac{2\pi\ell^{2}}{(r_{h}^{2}+\ell^{2})^{3/2}}\bigg|_{0}
           = \frac{2\pi}{\ell}.
\end{equation}
The second derivative vanishes identically, so the expansion contains
\emph{no quadratic term}.
Integrating term by term yields the exact small-horizon series
\begin{equation}
  \boxed{S(r_{h}) = 2\pi\ell\,r_{h}
  + \frac{\pi}{3\ell}\,r_{h}^{3}
  + \mathcal{O}(r_{h}^{5}/\ell^{3})}.
  \label{S-smallrh}
\end{equation}
The leading correction is \textit{linear} in $r_{h}$, in contrast
to the quadratic classical behavior $S_{\text{cl}}=\pi r_{h}^{2}$
and to the logarithmic corrections arising from GUP models.
This linear behavior reflects the geometry of the minimal core:
for $r_{h}\ll\ell$, the horizon is a thin annulus at areal radius
$R_{h}\approx\ell$, and its area grows as $4\pi\ell\cdot r_{h}$,
proportional to $r_{h}$.
The absence of a quadratic term follows from the even symmetry of
$R_h(r_h)=\sqrt{r_h^2+\ell^2}$ about $r_h=0$ and can also be
verified directly from the explicit integral~\eqref{Entropy}.
The next contribution enters at order $r_h^3/\ell$, a characteristic
signature of the smooth minimal-core geometry.

\subsubsection{Corrections for large horizons
\texorpdfstring{($r_{h}\gg\ell$)}{(r\_h >> ell)}.}
For $r_{h}\gg\ell$:
\begin{equation}
  S(r_{h}) \approx \pi r_{h}^{2}
  + \frac{\pi\ell^{2}}{2}
  + \pi\ell^{2}\ln\!\left(\frac{2r_{h}}{\ell}\right)
  + \mathcal{O}(\ell^{4}/r_{h}^{2}),
  \label{S-largerh}
\end{equation}
where the leading term is the classical area law and the subleading
corrections include a logarithm.

\begin{figure}[!htbp]
\centering
\includegraphics[width=\linewidth]{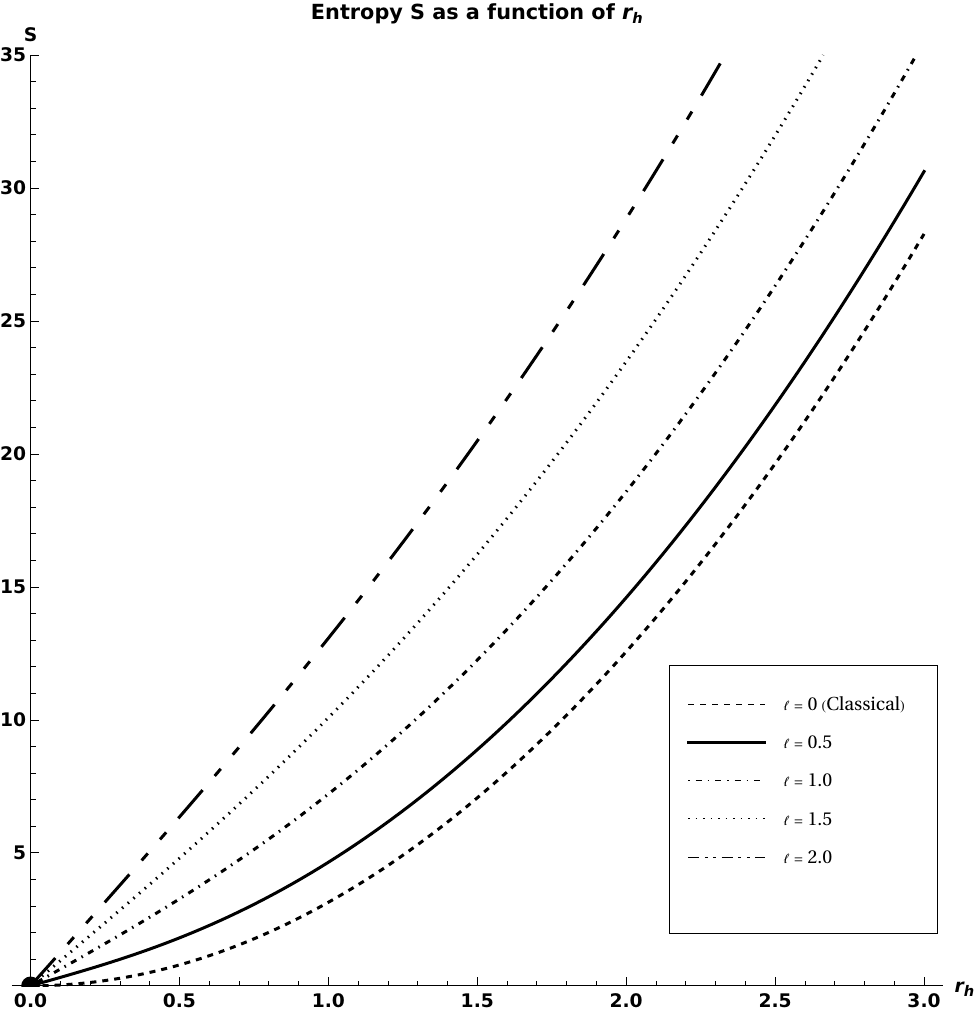}
\caption{Entropy $S$ as a function of horizon radius $r_{h}$
  for different values of $\ell$ (dimensionless units).
  The classical Bekenstein--Hawking entropy $S=\pi r_{h}^{2}$
  (dashed) is recovered for $\ell\to 0$.
  For $\ell>0$, the entropy vanishes linearly as
  $S\approx 2\pi\ell\,r_{h}$ for $r_{h}\ll\ell$
  (dash-dotted), and vanishes at $r_{h}=0$ automatically,
  consistent with the degenerate extremal endpoint.}
\label{fig:S}
\end{figure}

\section{Thermodynamic Stability and Phase Structure}
\label{sec4}

\subsection{Heat capacity and two-phase structure}

$C=dM/dT_H$ coincides with
Joshi and Joshi~\cite{joshi2026thermodynamic}
(cf.\ Table~\ref{tab:attribution}); we keep only the closed form
and the values needed below (intermediate steps in~\ref{app:C_derivation}):
\begin{equation}
    C = -\frac{8\pi M^{3}\sqrt{4M^{2}-\ell^{2}}}{2M^{2}-\ell^{2}}.
  \label{HeatCap}
\end{equation}
$C<0$ (\emph{Phase~I}, unstable) for $M>M^{*}=\ell/\sqrt2$ and
$C>0$ (\emph{Phase~II}, locally stable) for $\ell/2<M<M^{*}$, the
two phases separated by a Davies-type~\cite{davies1977thermodynamics}
divergence $C\to\pm\infty$ at $M^{*}$ -- second-order, since
$S(M^{*})=\pi\ell^2[\sqrt2+\ln(1+\sqrt2)]$ stays continuous there
(the divergence shares only the qualitative structure of, and
should not be confused with, the AdS Hawking--Page
transition~\cite{hawkingpage1983}). As $M\to M_{\min}=\ell/2$,
$C\to0^{+}$, the thermal analogue of the extremal
Reissner--Nordstr\"om endpoint. This divergence is visible
indirectly through the corresponding divergence of $R_{\rm GTD}$ in
Fig.~\ref{fig:RGTD} (Sec.~\ref{sec:GTD}); Phase~II is the phase
thermodynamically selected by the free energy of
Sec.~\ref{sec:freeenergy} below.

\subsection{Helmholtz free energy and canonical ensemble}
\label{sec:freeenergy}

To characterise the phase structure in the canonical ensemble
(fixed temperature $T$), we compute the Helmholtz free energy $F=M-T_HS$ explicitly, using
the first-law entropy of Eq.~\eqref{Entropy} -- as opposed to the
area-law-based free energy of~\cite[Eq.~(15)]{joshi2026thermodynamic}
(\S\ref{sec:entropy_SV_comparison}).
Using equations~\eqref{TH} and~\eqref{Entropy} with
$r_{h}=\sqrt{4M^{2}-\ell^{2}}$ and $R_{h}=2M$:
\begin{align}
  F &= M - T_{H}\,S \nonumber\\
  &= M - \frac{\sqrt{4M^{2}-\ell^{2}}}{16\pi M^{2}}
    \cdot\pi\!\left[r_{h}R_{h}
    + \ell^{2}\ln\!\frac{r_{h}+R_{h}}{\ell}\right] \nonumber\\
  &= \boxed{\frac{4M^{2}+\ell^{2}}{8M}
    - \frac{\ell^{2}\,r_{h}}{16M^{2}}
      \ln\!\frac{r_{h}+2M}{\ell}},
  \label{HelmholtzF}
\end{align}
where in the last step we used $r_{h}^{2}=4M^{2}-\ell^{2}$ to
simplify $M - r_{h}^{2}/(8M) = (4M^{2}+\ell^{2})/(8M)$.

\paragraph{Properties of $F$.}

\begin{enumerate}
  \item \textit{Differential identity.}
  Using the first law $dM=T_{H}\,dS$, one finds
  \begin{equation}
    \frac{\partial F}{\partial M}\bigg|_{\ell}
    = -S\,\frac{dT_{H}}{dM},
    \label{dFdM}
  \end{equation}
  so $F$ is stationary with respect to $M$ wherever
  $dT_{H}/dM=0$, i.e.\ precisely at $M=M^{*}$.
  This confirms that $M^{*}$ is a saddle point of $F(M)$
  at fixed $\ell$, consistent with the change of stability.

  \item \textit{Endpoint value.}
  At $M=M_{\min}=\ell/2$, $r_{h}=0$ and
  \begin{equation}
    F(M_{\min}) = \frac{4(\ell/2)^{2}+\ell^{2}}{4\ell} = \frac{\ell}{2}
    = M_{\min},
  \end{equation}
  so the free energy of the degenerate extremal remnant equals its
  rest mass, as expected for a zero-temperature, zero-entropy
  state with $F = M - 0\cdot 0 = M$.

  \item \textit{Canonical phase structure.}
  For $T<T_{H}^{\max}=1/(8\pi\ell)$, the equation $T_{H}(M)=T$
  has two solutions: $M_{1}\in(\ell/2,\,\ell/\sqrt{2})$
  (Phase~II, $C>0$) and $M_{2}>\ell/\sqrt{2}$ (Phase~I, $C<0$).
  The thermodynamically preferred phase minimises $F$ at the same
  temperature.
  Since $\partial^{2}F/\partial T^{2}=-\partial S/\partial T
  =-C/T_{H}<0$ in Phase~I and $>0$ in Phase~II, Phase~II
  (the stable phase with $C>0$) is the local free-energy
  minimum and is therefore thermodynamically selected.
  For $T>T_{H}^{\max}$ no equilibrium black hole solution exists
  in the canonical ensemble; the system evaporates completely to
  the degenerate extremal remnant.

  \item \textit{Continuity of $F$ and second-order character.}
  Both $F$ and $S$ are continuous functions of $M$ across $M^{*}$;
  only the second derivative of $F$ with respect to $T_{H}$ is
  discontinuous (via $C = -T_{H}\partial^{2}F/\partial T_{H}^{2}$
  which diverges). This confirms the second-order (continuous)
  character of the transition, in agreement with the Davies
  classification~\cite{davies1977thermodynamics}.
\end{enumerate}
\subsection{Geometrothermodynamics: a Legendre-invariant curvature
  diagnostic}
\label{sec:GTD}

The two-phase structure identified in Sec.~\ref{sec4} through
$C(M)$ and $F(M)$ has a purely geometric counterpart: the
equilibrium thermodynamics of the model can be encoded in a
Riemannian metric on the space of equilibrium states, whose
curvature is expected to diverge precisely where the heat capacity
does~\cite{ruppeiner1995riemannian,weinhold1975metric}. We construct
this metric using the Legendre-invariant formalism of
Quevedo~\cite{quevedo2007geometrothermodynamics,
quevedosanchez2009geometrothermodynamics}
(``geometrothermodynamics'', GTD), applied here for the first time
to $S(r_h,\ell)$ (cf.\ Table~\ref{tab:attribution}).

\subsubsection{Formalism}

We use Quevedo's Legendre-invariant metric
$G^{II}$~\cite{quevedo2007geometrothermodynamics,
quevedosanchez2009geometrothermodynamics}, the standard choice in the
applied black-hole literature. For a two-variable potential
$\Phi=\Phi(E^1,E^2)$ its pullback to the space of equilibrium states
reduces to
\begin{equation}
  ds_{G}^{2} = \bigl(E^{1}\Phi_{1}+E^{2}\Phi_{2}\bigr)
    \Bigl[-\Phi_{11}\,(dE^{1})^{2}+\Phi_{22}\,(dE^{2})^{2}\Bigr].
  \label{GTDmetric}
\end{equation}
where subscripts denote partial derivatives; for two extensive
variables and the standard choice $\eta=\mathrm{diag}(-1,1)$, the
off-diagonal contribution $\propto(\eta_{1}+\eta_{2})\Phi_{12}$
cancels identically, leaving the diagonal form above.

\subsubsection{Specialisation to the present model}

We take $E^{1}=S$, $E^{2}=\ell$, and $\Phi=M(S,\ell)$, treating the
minimal-length scale $\ell$ as a state parameter on the same
footing as the charge or the noncommutative parameter in analogous
constructions for Reissner--Nordstr\"om or noncommutative-inspired
black holes. Equation~\eqref{GTDmetric} becomes
\begin{equation}
  ds_{G}^{2} = \Lambda(S,\ell)
    \Bigl[-M_{SS}\,dS^{2}+M_{\ell\ell}\,d\ell^{2}\Bigr],
  \qquad
  \Lambda \equiv S\,M_{S} + \ell\,M_{\ell}.
  \label{Lambda_def}
\end{equation}
Because $S(r_{h})$ of Eq.~\eqref{Entropy} cannot be inverted for
$r_{h}(S)$ in elementary closed form, we evaluate $M_{S}$, $M_{SS}$,
$M_{\ell}$ and $M_{\ell\ell}$ by treating $(r_{h},\ell)$ as an
auxiliary parametrisation of the equilibrium manifold and applying
the implicit function theorem at fixed $S$ throughout; this
introduces no loss of rigour, since $S(r_{h},\ell)$ is a smooth
bijection of $r_{h}$ at fixed $\ell$ (Eq.~\eqref{dSdrh}:
$\partial S/\partial r_{h}=2\pi R_{h}>0$). The complete step-by-step
derivation of $M_{\ell}$ and $M_{\ell\ell}$ is given in~\ref{app:GTD_derivation}.

\paragraph{First derivatives.}
By definition of the first law, $M_{S}=T_{H}$ (Eq.~\eqref{TH}).
Applying the implicit function theorem
to $\partial S/\partial\ell$ at fixed $r_{h}$ and combining with the
chain rule (complete algebra in~\ref{app:GTD_derivation})
gives
\begin{equation}
  \boxed{
  M_{\ell} = \frac{\ell}{2R_{h}}
    \left[1-\frac{r_{h}}{R_{h}}\,
      \ln\!\left(\frac{r_{h}+R_{h}}{\ell}\right)\right]}.
  \label{Mell_closed}
\end{equation}
As a consistency check, along the boundary $r_{h}=0$ --- where
$S\equiv0$ identically for every $\ell$
(Sec.~\ref{sec:degenerate_horizon}) --- Eq.~\eqref{Mell_closed}
gives $M_{\ell}|_{r_{h}=0}=1/2$, exactly reproducing
$dM_{\min}/d\ell=d(\ell/2)/d\ell=1/2$: since $S=0$ along this entire
locus, the fixed-$S$ derivative $M_{\ell}$ must coincide there with
the ordinary derivative of $M_{\min}(\ell)=\ell/2$, which it does,
identically and for every $\ell$, not merely at a single point.

\paragraph{Second derivatives.}
For $M_{SS}$, the standard thermodynamic identity
$M_{SS}=\partial T_{H}/\partial S=T_{H}\,(dT_{H}/dM)$ (using
$dM/dS=T_{H}$) applies directly with $T_{H}(M)$ and $dT_{H}/dM$
(Eq.~\eqref{dTdM}), giving
\begin{equation}
  M_{SS} = T_{H}\,\frac{dT_{H}}{dM}
  = \frac{\ell^{2}-2M^{2}}{128\,\pi^{2}\,M^{5}}.
  \label{MSS_closed}
\end{equation}
This vanishes exactly at $M=M^{*}=\ell/\sqrt2$ --- the Davies point
already identified through $C(M)$ in Sec.~\ref{sec4} --- since
$M_{SS}=T_{H}/C$ and $C$ diverges there. Differentiating $M_{\ell}$
of Eq.~\eqref{Mell_closed} a second time by the same
implicit-function procedure (complete algebra in~\ref{app:GTD_derivation}) yields
\begin{equation}
  \boxed{
  M_{\ell\ell} = \frac{r_{h}^{2}}{R_{h}^{3}}
  + \frac{r_{h}(3\ell^{2}-r_{h}^{2})}{2R_{h}^{4}}\,
    \ln\!\left(\frac{r_h+R_h}{\ell}\right)
  + \frac{\ell^{2}(\ell^{2}-r_{h}^{2})}{2R_{h}^{5}}\,
    \left[\ln\!\left(\frac{r_h+R_h}{\ell}\right)\right]^{2}
  }.
  \label{Mellell_closed}
\end{equation}
The same consistency check applies: at $r_{h}=0$,
Eq.~\eqref{Mellell_closed} gives $M_{\ell\ell}=0$ identically for
every $\ell$, exactly reproducing
$d^{2}M_{\min}/d\ell^{2}=d^{2}(\ell/2)/d\ell^{2}=0$.

\subsubsection{Metric degeneracy and thermodynamic curvature}

Equations~\eqref{Lambda_def}--\eqref{Mellell_closed} give the GTD
line element in fully closed form, parametrised by $(r_{h},\ell)$.
Two loci are singled out analytically.

\begin{itemize}
  \item \textit{Davies point $M=M^{*}=\ell/\sqrt2$ ($r_{h}=\ell$).}
    Here $M_{SS}=0$ exactly (Eq.~\eqref{MSS_closed}), while
    \begin{equation}
      \Lambda(M^{*}) = \frac{\ell}{8}\bigl[3\sqrt2-\ln(1+\sqrt2)\bigr]
        \approx 0.420\,\ell,
      \qquad
      M_{\ell\ell}(M^{*}) = \frac{\sqrt2+\ln(1+\sqrt2)}{4\ell}
        \approx \frac{0.574}{\ell}
      \label{GTD_atMstar}
    \end{equation}
    (obtained by substituting $r_{h}=\ell$ into
Eq.~\eqref{Mell_closed}, using $S(M^{*})$ and $T_{H}^{\max}$
from Sec.~\ref{sec4}) are both finite and nonzero.
    Hence $g_{SS}\to0$ while $g_{\ell\ell}$ stays finite: the
    metric degenerates in the entropy direction exactly where
    $C(M)$ diverges.
  \item \textit{Degenerate extremal endpoint $M=M_{\min}=\ell/2$
    ($r_{h}=0$).} Here instead $M_{\ell\ell}=0$ exactly, while
    \begin{equation}
      \Lambda(M_{\min}) = \frac{\ell}{2},
      \qquad
      M_{SS}(M_{\min}) = \frac{1}{8\pi^{2}\ell^{3}}
      \label{GTD_atMmin}
    \end{equation}
    are finite and nonzero. Hence $g_{\ell\ell}\to0$ while $g_{SS}$
    stays finite: the metric degenerates in the $\ell$-direction,
    producing a second, independent curvature singularity, located
    not at a phase transition of $C(M)$ but exactly at the
    degenerate extremal remnant of Sec.~\ref{sec:degenerate_horizon}.
\end{itemize}

Because the two degeneracies occur in orthogonal directions of the
diagonal metric~\eqref{Lambda_def}, they are structurally distinct:
the Davies-point degeneracy is the standard GTD signature of a
heat-capacity divergence~\cite{quevedosanchez2009geometrothermodynamics},
while the remnant-point degeneracy has, to our knowledge, no
counterpart in the GTD literature on regular black holes and is
specific to the presence of a genuine (degenerate) horizon at
$S=0$ --- a feature absent from every other minimal-length model in
Table~\ref{tab:comparison}, where evaporation either continues
indefinitely or halts at a horizonless core with $S>0$.

\begin{figure}[!htbp]
\centering
\includegraphics[width=\linewidth]{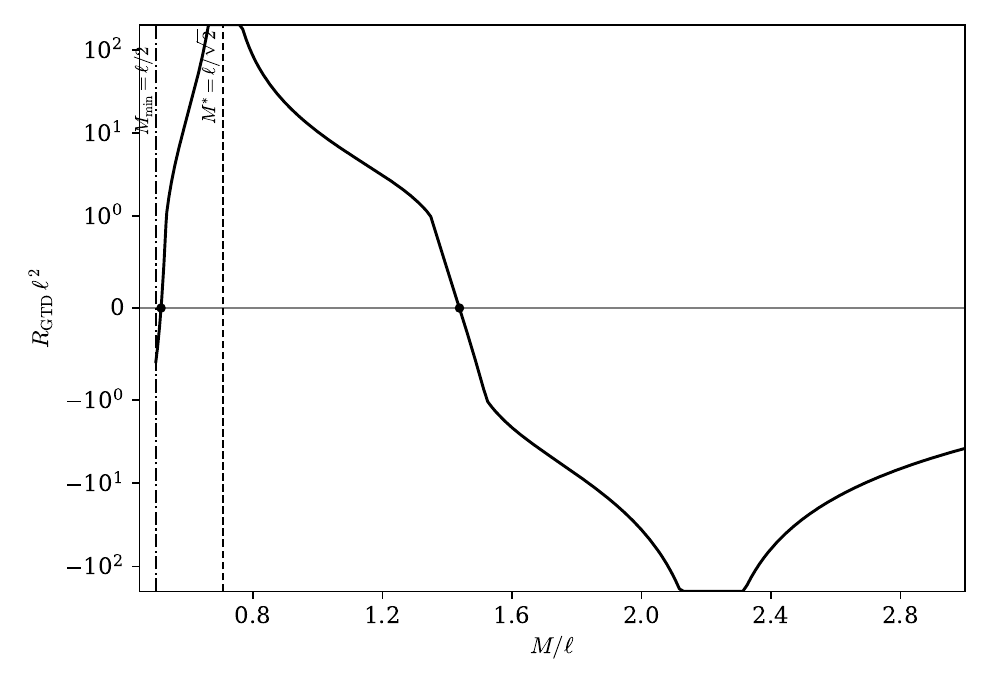}
\caption{Thermodynamic curvature scalar $R_{\rm GTD}$
  (Eq.~\eqref{RGTD_formula}) as a function of $M/\ell$. Two
  divergences are visible: at $M^{*}=\ell/\sqrt2$ (vertical dashed
  line), coinciding with the Davies-type heat-capacity transition
  of Eq.~\eqref{HeatCap} (Sec.~\ref{sec4}), and at $M_{\min}=\ell/2$
  (vertical dot-dashed line), coinciding with the degenerate extremal
  remnant of Sec.~\ref{sec:degenerate_horizon}. The sign of
  $R_{\rm GTD}$ differs on the two sides of $M^{*}$, tracking the
  sign change of $C(M)$.}
\label{fig:RGTD}
\end{figure}

The full thermodynamic curvature follows from the standard formula
for an orthogonal two-dimensional metric $ds^{2}=g_{SS}\,dS^{2}
+g_{\ell\ell}\,d\ell^{2}$,
\begin{equation}
  R_{\rm GTD} = -\frac{1}{\sqrt{g_{SS}g_{\ell\ell}}}
  \left[
    \frac{\partial}{\partial S}\!\left(
      \frac{\partial_{S}g_{\ell\ell}}{\sqrt{g_{SS}g_{\ell\ell}}}
    \right)
    +
    \frac{\partial}{\partial \ell}\!\left(
      \frac{\partial_{\ell}g_{SS}}{\sqrt{g_{SS}g_{\ell\ell}}}
    \right)
  \right],
  \label{RGTD_formula}
\end{equation}
evaluated with $g_{SS}=-\Lambda M_{SS}$,
$g_{\ell\ell}=\Lambda M_{\ell\ell}$ from
Eqs.~\eqref{Lambda_def}--\eqref{Mellell_closed}. Because
$\partial_{\ell}g_{SS}$ requires a further fixed-$S$ derivative of
an already lengthy expression, we evaluate
Eq.~\eqref{RGTD_formula} symbolically rather than displaying the
resulting multi-term expression;~\ref{app:GTD_CAS} reports
the algorithm used, quantifies why the fully expanded scalar is not
displayable in print, and gives the independent computer-algebra
verification -- performed from scratch in \textsc{SymPy}, starting
only from $S(r_h,\ell)$ and $M=R_h/2$ -- of every closed-form
quantity in this subsection, including $\Lambda(M^{*})$ and
$M_{\ell\ell}(M^{*})$ to fifteen significant figures. Fig.~\ref{fig:RGTD} shows the
result as a function of $M/\ell$, confirming the two divergences
identified analytically above and their opposite sign (reflecting
the change of stability across $M^{*}$, consistently with the sign
change of $C$ established in Sec.~\ref{sec4}).

\paragraph{Physical reading.}
In the Ruppeiner interpretation~\cite{ruppeiner1995riemannian},
$|R_{\rm GTD}|$ is a proxy for the correlation volume of the
underlying microscopic degrees of freedom, and its divergence
signals a change in the effective interaction between them. The
divergence at $M^{*}$ is the expected counterpart, in thermodynamic
geometry, of the Davies-type transition already established through
$C(M)$ and $F(M)$ (Sec.~\ref{sec4}). The divergence at
$M_{\min}$ is new: it suggests that the approach to the degenerate
extremal remnant is accompanied by a divergence of the same
correlation-length proxy, i.e.\ that the remnant is approached
through a regime of maximally correlated microstructure rather than
a smooth freeze-out --- a statement that remains at the level of
semiclassical thermodynamic geometry and does not, by itself,
constitute a microscopic derivation (cf.\ the discussion of quantum
stability in Sec.~\ref{sec5}).
\subsection{Energy conditions and nature of the effective source}
\label{sec:energy}

Using the components~\eqref{EMT}, the standard energy conditions
for the anisotropic fluid take the following forms.

\textit{Null energy condition (NEC)}, radial sector
$\rho+p_{r}\geq 0$:
\begin{equation}
  \rho + p_{r} = \frac{\ell^{2}(2M-R)}{4\pi R^{5}}.
  \label{NECr}
\end{equation}
This is \emph{negative} for $R>2M$ (exterior region) and positive
for $R<2M$ (interior region).
The radial NEC is therefore \emph{violated outside the horizon}
and satisfied inside.
This pattern is typical of regular black hole models
(cf.~Barcel\'o and Visser~\cite{barcelo2000scalar}): the radial
null energy condition necessarily fails in at least part of the
exterior in any geometry that resolves the central singularity
while maintaining asymptotic flatness.
This violation has a direct thermodynamic consequence
(\S\ref{sec:BC}): it prevents the first-law entropy of
Eq.~\eqref{Entropy} from coinciding with
the bare area law $S=A/4=\pi R_h^2$, in line with the general
pattern established for other NEC-violating regular black
holes~\cite{maZhao2014correctedfirstlaw}.

\textit{NEC, tangential sector} $\rho+p_{\vartheta}\geq 0$:
\begin{equation}
  \rho + p_{\vartheta} = \frac{3M\ell^{2}}{8\pi R^{5}} > 0,
  \label{NECt}
\end{equation}
\emph{satisfied everywhere} for $M>0$.
This result follows directly from the effective
stress-energy components~\eqref{EMT}:
\begin{equation}
  \rho + p_{\vartheta}
  = \frac{\ell^{2}(4M-R)}{8\pi R^{5}}
  + \frac{\ell^{2}(R-M)}{8\pi R^{5}}
  = \frac{3M\ell^{2}}{8\pi R^{5}} > 0.
\end{equation}
The tangential pressure $p_\vartheta = \ell^2(R-M)/(8\pi R^5)$
is positive throughout the exterior ($R>M$), and the combination
$\rho+p_\vartheta$ is strictly positive at all areal radii,
reflecting the mild geometric repulsion in the angular directions
that maintains the regularity of the 2-sphere fibration as $r\to 0$.
Unlike many regular black hole models where tangential NEC
violations are generic~\cite{barcelo2000scalar}, the geometric
minimal-length deformation \emph{satisfies} the tangential NEC
globally, a property of the specific areal-radius deformation $R(r)=\sqrt{r^2+\ell^2}$.
The magnitude $3M\ell^{2}/(8\pi R^{5})$ falls off as $\ell^{2}/R^{5}$
and is negligibly small at any astrophysical scale $R\gg\ell$.

\textit{Strong energy condition (SEC)}:
\begin{equation}
  \rho + p_{r} + 2p_{\vartheta}
    = \frac{M\ell^{2}}{4\pi R^{5}} > 0.
\end{equation}
The trace part of the SEC is \emph{satisfied everywhere} for $M>0$.
The detailed computation reads
\begin{equation}
  \rho + p_{r} + 2p_{\vartheta}
  = \frac{\ell^{2}(4M-R)}{8\pi R^{5}}
  - \frac{\ell^{2}}{8\pi R^{4}}
  + \frac{2\ell^{2}(R-M)}{8\pi R^{5}}
  = \frac{2M\ell^{2}}{8\pi R^{5}}
  = \frac{M\ell^{2}}{4\pi R^{5}} > 0.
\end{equation}
However, the \emph{full} SEC requires in particular that
$\rho+p_{r}\geq 0$ (the radial NEC), which fails in the exterior
$R>2M$ as established above.
The SEC is therefore globally \emph{violated in the exterior region}
via the failure of the radial NEC, while the trace condition
$\rho+p_r+2p_\vartheta = M\ell^{2}/(4\pi R^{5})>0$ is satisfied
throughout the entire spacetime.
This nuanced pattern---trace SEC satisfied everywhere, full SEC
violated only in the exterior via the radial NEC, and tangential
NEC satisfied everywhere---is a feature of the geometric minimal-length deformation compared with other regular
black hole models~\cite{hayward2006formation,Nicolini2006} where
tangential NEC violations are typical.

The magnitude of the radial NEC violation scales as $\ell^{2}/R^{5}$
(cf.~Eq.~\eqref{NECr}), falling off rapidly outside the core.
At astrophysical scales $R\gg\ell$, the effective source is entirely
negligible and the geometry is indistinguishable from Schwarzschild.
In Planck units with $\ell\sim\ell_{P}$, the violation is of
order unity at the core, which is the natural regime where classical
general relativity is expected to break down anyway.

\section{Observational Signatures: Shadow and Gravitational Lensing}
\label{sec:observational}

\subsection{Photon sphere and black hole shadow}
\label{sec:shadow}

Imposing the photon-sphere condition $dV_{\rm eff}/dr=0$ on
$V_{\rm eff}(r)=f(R)/R^2$ gives (full derivation in~\ref{app:photon_derivation})
\begin{equation}
  R = 3M,
  \label{ph-R}
\end{equation}
with coordinate radius
\begin{equation}
  r_{\rm ph} = \sqrt{R_{\rm ph}^{2} - \ell^{2}} = \sqrt{9M^{2} - \ell^{2}},
  \label{rph}
\end{equation}
real for every black hole in this model since $M\geq\ell/2>\ell/3$,
and critical impact parameter
\begin{equation}
  b_{c} = \frac{R_{\rm ph}}{\sqrt{f(R_{\rm ph})}} = 3\sqrt{3}\,M.
  \label{bc}
\end{equation}
Both $R_{\rm ph}=3M$ and $b_{c}=3\sqrt3\,M$ are \emph{independent of
$\ell$}, matching Schwarzschild and Tsukamoto's Simpson--Visser
result~\cite{tsukamoto2021gravitational} (cf.\ Table~\ref{tab:attribution});
only the coordinate radius $r_{\rm ph}$ retains an explicit
$\ell$-dependence, which
resurfaces in the strong-field lensing coefficients of
Section~\ref{sec:strongfield}.

\subsection{Gravitational lensing}
\label{sec:lensing}

The exact deflection angle follows from the orbit equation with
impact parameter $b=R_{0}/\sqrt{f(R_{0})}$ set by the turning-point
condition $\dot r=0$ at $r=r_0$; changing variable to
$R=\sqrt{r^{2}+\ell^{2}}$ gives, after straightforward algebra,
\begin{equation}
  \alpha(R_{0}) = 2\int_{R_{0}}^{\infty}
    \frac{R_{0}\,dR}{\sqrt{R^{2}-\ell^{2}}\;\sqrt{f(R_{0})\,R^{2}
    - f(R)\,R_{0}^{2}}}
    - \pi.
  \label{alpha_exact}
\end{equation}
This is, up to notation, exactly Tsukamoto's~\cite{tsukamoto2021gravitational}
integral in his standard radial coordinate $\rho=R$ (his
Eq.~(2.22)); we keep the independent $r$-based route because it is
reused in Sec.~\ref{sec:strongfield} to re-derive the coefficient
$a$ by a different method (a direct quadratic expansion at the
photon sphere, rather than Bozza's compactified variable). An equivalent form used below is
\begin{equation}
  \alpha(R_{0}) = 2\int_{R_{0}}^{\infty}
    \frac{dR}{\sqrt{f(R)}\,\sqrt{R^{2}-\ell^{2}}}
    \left[\frac{f(R_{0})}{f(R)}\frac{R^{2}}{R_{0}^{2}} - 1\right]^{-1/2}
    - \pi\,,
  \label{alpha_equiv}
\end{equation}
which reduces to the standard Schwarzschild integral at $\ell=0$.

\subsubsection{Weak-field limit and analytical minimal-length correction}
\label{sec:weakfield}

Expanding the integrand of~\eqref{alpha_exact} for $R_{0}\gg M,\ell$
reproduces the standard Einstein deflection $\alpha\simeq4M/b$ at
leading order, with $b\simeq R_0$; the minimal-length deformation
adds an independent correction from the factor
$1/\sqrt{R^2-\ell^2}$, evaluated exactly in~\ref{app:weakfield_derivation}. The full weak-field
deflection angle to leading order in both $M/b$ and $\ell^2/b^2$ is
\begin{equation}
  \boxed{
    \alpha \simeq \frac{4M}{b}
    + \frac{\pi\ell^{2}}{4b^{2}}
    + \mathcal{O}\!\left(\frac{M^{2}}{b^{2}},\,\frac{M\ell^{2}}{b^{3}}\right).
  }
  \label{weak_corrected}
\end{equation}
Setting $M=0$ reproduces the massless Simpson--Visser weak-field
deflection $\alpha\simeq\pi\ell^2/(4b^2)$ of Nascimento et
al.~\cite{nascimento2020weak} and \"Ovg\"un~\cite{ovgun2020weak}
(cf.\ Table~\ref{tab:attribution}); what is new here is the explicit
combination with the $4M/b$ Einstein term at the same order for
$M\neq0$.

The correction is independent of $M$, positive (the deformation
marginally enhances light bending), and astrophysically negligible:
for $\ell\sim\ell_P$ the ratio to the Einstein term,
$\pi\ell^2/(16Mb)$, is of order $\ell_P^2/(M\,r_{\rm Schw})
\lesssim10^{-70}$, so galaxy- and cluster-scale lensing remains
indistinguishable from Schwarzschild.

\subsubsection{Strong-field limit and photon ring}
\label{sec:strongfield}

As $R_{0}\to R_{\rm ph}=3M$ the deflection angle diverges
logarithmically.
We extract the analytic behaviour via the strong-field expansion
of Bozza~\cite{bozza2002}.
The key step is to establish that the denominator in the integrand
of~\eqref{alpha_exact} behaves quadratically near the photon sphere,
with a coefficient derived below.

\paragraph{Extraction of the strong-lensing coefficient $a$.}

A quadratic expansion of $f(R)$ at $R_{\rm ph}=3M$ gives
$f(R_0)R^2-f(R)R_0^2\simeq\mathcal{C}[(R-R_{\rm ph})^2-(R_0-R_{\rm
ph})^2]$ with $\mathcal{C}\equiv f_{\rm ph}-\tfrac12f''_{\rm
ph}R_{\rm ph}^2=1$ exactly; combined with the expansion of
$\sqrt{R^2-\ell^2}$ near $R_{\rm ph}$ and
$b-b_c=(\sqrt3/2M)u_0^2$ (turning-point offset $u_0=R_0-R_{\rm
ph}$), this isolates the divergent part of
Eq.~\eqref{alpha_exact} as $\alpha_{\rm div}=-a\ln(b/b_c-1)+{\rm
const}$ -- full algebra in~\ref{app:a_derivation} -- with
\begin{equation}
  a = \frac{R_{\rm ph}}{\sqrt{R_{\rm ph}^{2}-\ell^{2}}}
    = \frac{3M}{\sqrt{9M^{2}-\ell^{2}}}
    = \frac{3}{\sqrt{\,9 - \ell^{2}/M^{2}\,}}.
  \label{coeff_a}
\end{equation}
$a=1$ at $\ell=0$ (Schwarzschild~\cite{bozza2002}) and $a>1$ for
$\ell>0$; for $M\gg\ell$,
\begin{equation}
  a \simeq 1 + \frac{\ell^{2}}{18M^{2}} + \mathcal{O}(\ell^{4}/M^{4}).
  \label{approx_a}
\end{equation}
This closed form matches Tsukamoto's coefficient
$\bar a$~\cite{tsukamoto2021gravitational} (his Eq.~(3.22); cf.\
Table~\ref{tab:attribution}).

\paragraph{The regular coefficient $\bar{b}$.}
\label{par:barb_correction}
A naive subtraction of the singular part of
Eq.~\eqref{alpha_equiv} directly in the bare $R$-variable diverges
as $R\to\infty$, so we adopt Tsukamoto's convergent,
compactified-variable expression~\cite{tsukamoto2021gravitational}
rather than re-deriving it from Bozza's
method~\cite{bozza2002}; the divergence and the full construction
are given in~\ref{app:a_derivation}. Evaluating his
expression by Gaussian quadrature, with $R(r)=\sqrt{r^2+\ell^2}$
identified with his standard radial coordinate $\rho$, gives
\begin{equation}
  \bar b = a\ln(6M) + I_R - \pi,
  \label{barb_correct}
\end{equation}
which reproduces Tsukamoto's tabulated values at $\ell/M=1,1.5$ to
five significant figures (Table~\ref{tab:lensing}) and reduces, at
$\ell=0$ ($a=1$), to the known analytic Schwarzschild
result~\cite{bozza2002}
\begin{equation}
  \bar{b}\big|_{\ell=0}
  = \ln\!\bigl(216(7-4\sqrt{3})\bigr) - \pi \approx -0.4002.
  \label{bbar_Schw}
\end{equation}

\begin{table}[!htbp]
\centering
\caption{Strong-field lensing coefficients as a function of $\ell/M$
  (with $\ell<\ell_{\rm max}=3M$ for the photon sphere to exist for
  any black hole). The coefficient $a$ is given exactly by
  Eq.~\eqref{coeff_a} and coincides with $\bar a$
  of~\cite{tsukamoto2021gravitational}; $\bar{b}$ is computed from
  the convergent expression~\eqref{barb_correct}, following
  Tsukamoto's regularisation, and validated against his
  published Table~I at $\ell/M=1$ and $1.5$ (agreement to five
  significant figures); $r_n=e^{-2\pi/a}$ is the flux ratio between
  consecutive photon-ring images.}
\label{tab:lensing}
\renewcommand{\arraystretch}{1.3}
\begin{tabular}{cccc}
\hline
$\ell/M$ & $a$ & $\bar{b}$ & $r_n = e^{-2\pi/a}$\\
\hline
$0$   & $1.0000$ & $-0.40023$ & $1.8674\times10^{-3}$\\
$0.5$ & $1.0142$ & $-0.41408$ & $2.04\times10^{-3}$\\
$1.0$ & $1.0607$ & $-0.46474$ & $2.68\times10^{-3}$\\
$1.5$ & $1.1547$ & $-0.59088$ & $4.33\times10^{-3}$\\
\hline
\end{tabular}
\par\vspace{4pt}
\begin{minipage}{0.92\linewidth}
  \footnotesize
  The entries for $\ell/M=1$ and $1.5$ reproduce
  Tsukamoto~\cite{tsukamoto2021gravitational}, Table~I
  ($a_{\rm SV}=m_*,\,1.5m_*$) to five significant figures. The entry for
  $\ell/M=0$ is the exact analytical result~\eqref{bbar_Schw}. The
  entry for $\ell/M=0.5$ is new (not tabulated
  in~\cite{tsukamoto2021gravitational}), obtained from the same
  convergent method. Angular positions $\theta_n$ of individual
  photon-ring images (Eq.~\eqref{theta_n}) require $\bar{b}$ in
  addition to $a$ and $b_c$; the flux ratio $r_n$ depends only on
  $a$ and is given analytically.
\end{minipage}
\end{table}

\paragraph{Photon-ring observables.}

The physical implications of $a$ and $\bar{b}$ are
directly linked to measurable photon-ring properties.
For a source and observer on the same side of the lens, the
$n$-th relativistic image appears at angular position
\begin{equation}
  \theta_{n} \approx \theta_{\infty}
    + \frac{b_{c}}{D_{\rm ol}}\,
      \exp\!\left(\frac{\bar{b} - 2\pi n}{a}\right),
  \label{theta_n}
\end{equation}
where $\theta_{\infty}=b_{c}/D_{\rm ol}$ is the shadow boundary
and $D_{\rm ol}$ is the observer--lens distance.
Equation~\eqref{theta_n} requires $\bar{b}$ from
Table~\ref{tab:lensing} for absolute positions.
The flux ratio between consecutive images depends only on $a$
and is given analytically:
\begin{equation}
  r_{n} = \frac{\mu_{n+1}}{\mu_{n}} = e^{-2\pi/a}.
  \label{rn}
\end{equation}
This ratio is therefore an exact, \emph{numerics-free} prediction
of the model, depending only on $a$ and hence, by the correspondence
above, identical to the prediction one would extract from
Tsukamoto's coefficient $\bar a$~\cite{tsukamoto2021gravitational}.

To illustrate the magnitude of the effect, we consider
$\ell = M$ (chosen large enough to make the deviation visible;
at astrophysical scales $\ell \ll M$ and all corrections are
suppressed by $(\ell/M)^{2}$, see Sec.~\ref{sec:obs_constraints}).
From Eq.~\eqref{coeff_a}:
\begin{equation}
  a\big|_{\ell=M} = \frac{3}{\sqrt{9-1}}
    = \frac{3}{\sqrt{8}} = \frac{3\sqrt{2}}{4} \approx 1.0607,
  \label{a_numerical}
\end{equation}
and the flux ratio from Eq.~\eqref{rn}:
\begin{equation}
  r_{n}\big|_{\ell=M} = e^{-2\pi/1.0607} \approx 2.676\times10^{-3},
  \label{rn_numerical}
\end{equation}
compared with the Schwarzschild value
$r_{n}^{\rm Schw} = e^{-2\pi} \approx 1.8674\times10^{-3}$.
The relative enhancement is approximately $43\%$, consistent
with Table~\ref{tab:lensing}.

The angular position of the first relativistic image follows
from Eq.~\eqref{theta_n} with $\bar{b}|_{\ell=M}\approx -0.46474$
(Table~\ref{tab:lensing}):
\begin{equation}
  \theta_1 \approx \theta_\infty
    + \frac{b_c}{D_{\rm ol}}
      \exp\!\left(\frac{-0.46474-2\pi}{1.0607}\right),
  \label{theta1_num}
\end{equation}
and the angular separation between the first and second images is
\begin{equation}
  \Delta\theta_{1,2} \approx
    \frac{b_{c}}{D_{\rm ol}}\,e^{(\bar{b}-2\pi)/a}
    \!\left(1 - e^{-2\pi/a}\right).
  \label{Delta_theta}
\end{equation}
Numerical values of $\Delta\theta_{1,2}$ in physical units
(microarcseconds) are obtained by inserting $\bar{b}$ from
Table~\ref{tab:lensing} and the measured mass and distance for
specific astrophysical targets.

\subsubsection{Shadow degeneracy versus photon-ring enhancement}
\label{sec:discriminant}

The observational profile of the model has a bipartite structure
(cf.\ Table~\ref{tab:attribution}), which we connect explicitly to
the degenerate-extremal-remnant scenario of
Section~\ref{sec:degenerate_horizon}.

\paragraph{Shadow: zero sensitivity to $\ell$.}
The critical impact parameter $b_c = 3\sqrt{3}M$ is exact and
independent of $\ell$ at all orders in $\ell/M$
(Eq.~\eqref{bc}).
The shadow angular diameter observed at distance $D$,
\begin{equation}
  \theta_{\rm sh} = \frac{2b_c}{D} = \frac{6\sqrt{3}\,M}{D},
\end{equation}
is exactly the Schwarzschild value.
Current Event Horizon Telescope measurements of M87*
($\theta_{\rm sh} = 42\pm3\,\mu{\rm as}$~\cite{EHT2019}) and
Sgr~A* are fully consistent with this prediction and impose
\emph{no constraint on $\ell$}.

\paragraph{Photon ring: analytic sensitivity to $\ell$.}
The coefficient $a$ controls the angular spacing and flux ratio
of successive photon-ring sub-images via
Eqs.~\eqref{theta_n}--\eqref{rn}.
From Eq.~\eqref{approx_a}, a fractional enhancement
$a - 1 = \epsilon$ requires
\begin{equation}
  \frac{\ell}{M} \approx \sqrt{18\epsilon}
    \simeq 4.24\,\sqrt{\epsilon}.
  \label{ell_constraint}
\end{equation}
As an illustration, a $1\%$ enhancement ($\epsilon=0.01$)
corresponds to $\ell/M \approx 0.42$, or a $0.1\%$ enhancement
to $\ell/M \approx 0.13$.
The flux ratio $r_n=e^{-2\pi/a}$ (Eq.~\eqref{rn}) is the
cleanest discriminant: it depends only on $a$ and not on
$\bar{b}$, making it an exact, numerics-free prediction.
Table~\ref{tab:lensing} shows that even at $\ell/M=0.5$,
the flux ratio deviates from the Schwarzschild value by
$\sim 9\%$, rising to $\sim 43\%$ at $\ell/M=1$.

\paragraph{Scaling with target mass.}
For M87* ($M\approx6.5\times10^{9}\,M_{\odot}$,
$D\approx16.8$\,Mpc), the angular shadow diameter is determined
by $M/D$ and is degenerate with Schwarzschild.
The photon-ring sub-image separation
$\Delta\theta_{1,2}\propto e^{-2\pi/a}/D_{\rm ol}$ is
enhanced relative to Schwarzschild by the factor
$e^{-2\pi(1/a-1)} = e^{-2\pi\epsilon/(1+\epsilon)}
\approx 1 + 2\pi\epsilon$ for small $\epsilon$, or
equivalently $(e^{-2\pi/a} - e^{-2\pi})/e^{-2\pi}
\approx 2\pi\ell^{2}/(18M^{2})$ for $\ell\ll M$.
For Sgr~A* ($M\approx4\times10^{6}\,M_{\odot}$,
$D\approx8.15$\,kpc), the shadow angular diameter is
$\theta_{\rm sh}\approx50\,\mu{\rm as}$ (exactly Schwarzschild),
while the photon-ring observables depend on $\ell/M$ through
the same coefficient $a$, following the same strategy
Tsukamoto~\cite{tsukamoto2021gravitational} used to estimate
$\ell/M$ for M87* from the EHT photon-ring diameter.

\section{Discussion and Conclusion}
\label{sec5}

\subsection{Degenerate extremal regular remnant and endpoint of
evaporation}

The endpoint (\S\S\ref{sec:degenerate_horizon}--\ref{sec:curv};
values in the Introduction and Table~\ref{tab:comparison}) is
qualitatively distinct from the two
outcomes usually discussed in the minimal-length literature: a
static horizonless core reached while $T_H$ and $C$ remain finite
and positive (Nicolini-type or Hayward-type models,
Table~\ref{tab:comparison}), or an evaporation with no natural
endpoint at all. Here the horizon itself degenerates onto the
regular center rather than shrinking to zero area or disappearing,
so the remnant retains a genuine, if degenerate, causal horizon
(Sec.~\ref{sec:penrose}) even though its temperature and entropy
both vanish. This has two consequences pursued below: because
$S\to0$ together with $T_H\to0$, there is no residual thermodynamic
driving force for further mass loss, unlike remnant scenarios that
retain nonzero equilibrium entropy; and because a genuine
(degenerate) horizon survives rather than disappearing, its
classical and quantum stability -- addressed in
Sec.~\ref{sec:modestability} and below -- is a well-posed question
in a way it would not be for a horizonless core.
\subsubsection{Linear mode stability against scalar perturbations}
\label{sec:modestability}

Before addressing the more speculative question of non-perturbative
quantum stability below, we first establish a modest but rigorous
classical result: the degenerate extremal remnant is linearly
mode-stable against massless scalar test-field perturbations. We
denote the multipole number by $L=0,1,2,\ldots$ throughout this
section, reserving $\ell$ exclusively for the minimal-length
parameter as elsewhere in this paper.

A minimally coupled massless scalar $\Psi$ obeying $\Box\Psi=0$ on
background~\eqref{metric}, decomposed as
$\Psi=\psi(r)R(r)^{-1}e^{-i\omega t}Y_{Lm}(\vartheta,\phi)$ and
written in the tortoise coordinate $r_{*}$ defined by
$dr_{*}/dr=1/f$, reduces by the standard procedure to a
Schr\"odinger-type radial equation of Regge--Wheeler
type~\cite{reggewheeler1957},
\begin{equation}
  \frac{d^{2}\psi}{dr_{*}^{2}} + \left[\omega^{2}-V_{L}(r)\right]\psi=0,
  \qquad
  V_{L}(r) = \frac{f}{R}\,\frac{d}{dr}\bigl(fR'\bigr)
    + \frac{f\,L(L+1)}{R^{2}}.
  \label{RWeq}
\end{equation}
Setting $R=r$ recovers the standard Schwarzschild scalar potential
$V_{L}=f\bigl[L(L+1)/r^{2}+2M/r^{3}\bigr]$, confirming
Eq.~\eqref{RWeq}.

Using $R'=r/R$, $R''=\ell^{2}/R^{3}$ (\ref{app:Kretschmann})
and $f_{r}=2Mr/R^{3}$, so that $d(fR')/dr=f_{r}R'+fR''
=2Mr^{2}/R^{4}+f\ell^{2}/R^{3}$, Eq.~\eqref{RWeq} evaluates in
closed form to
\begin{equation}
  \boxed{
  V_{L}(r) = \frac{2M f\,r^{2}}{R^{5}}
    + \frac{f^{2}\ell^{2}}{R^{4}}
    + \frac{f\,L(L+1)}{R^{2}}
  }.
  \label{Veff_closed}
\end{equation}

Each of the three terms in Eq.~\eqref{Veff_closed} is manifestly
non-negative wherever $f\geq0$: the first and third are products of
$f$ with manifestly non-negative factors, and the second is a
perfect square times $f^{2}\geq0$ regardless of the sign of $f$.
By the monotonicity of $f\circ R$ established in
Eq.~\eqref{fmonotone} (Sec.~\ref{sec:degenerate_horizon}), $f\geq0$
holds precisely on $r\geq r_{h}$, i.e.\ throughout the static
exterior, for \emph{any} $M\geq M_{\min}$ in this model. Consequently
$V_{L}(r)\geq0$ on the entire static exterior for the full physical
black-hole branch, not merely near the remnant.

\paragraph{Specialisation to the remnant.}
At the degenerate extremal endpoint $M=M_{\min}=\ell/2$, the horizon
coincides with the centre, $r_{h}=0$
(Sec.~\ref{sec:degenerate_horizon}), so $f\geq0$ --- and hence
$V_{L}\geq0$ by Eq.~\eqref{Veff_closed} --- holds on the
\emph{entire} domain $r\in[0,+\infty)$, with equality only at
$r=0$, where $f=0$. Using the near-centre expansion
$f\approx r^{2}/(2\ell^{2})$ of Eq.~\eqref{f_quadratic}, the leading
behaviour of Eq.~\eqref{Veff_closed} at $M=M_{\min}$ is
\begin{equation}
  V_{L}(r)\Big|_{M_{\min}}
  \;\xrightarrow{\,r\to0\,}\;
  \frac{L(L+1)}{2\ell^{4}}\,r^{2}
  + \frac{3}{4\ell^{6}}\,r^{4}
  + \mathcal{O}(r^{6}),
  \label{Vsmallr}
\end{equation}
i.e.\ quadratic vanishing for $L\geq1$ and quartic vanishing for the
monopole $L=0$, both with strictly positive coefficients: the
potential is smooth, non-negative, and vanishes at the degenerate
horizon at exactly the same order as $f$ itself.

\paragraph{Stability conclusion.}
For a mode of the form $\omega=i\kappa$ ($\kappa$ real), multiplying
Eq.~\eqref{RWeq} by $\psi^{*}$ and integrating over $r_{*}$ gives,
for any normalisable solution with $\psi\to0$ at both ends,
\begin{equation}
  \int\Bigl(|\dot\psi|^{2}+V_{L}|\psi|^{2}\Bigr)dr_{*}
  = -\kappa^{2}\int|\psi|^{2}\,dr_{*}.
\end{equation}
Since $V_{L}\geq0$ throughout the integration range, the left-hand
side is non-negative while the right-hand side is non-positive,
forcing $\kappa=0$: this is the standard energy argument used to
rule out exponentially growing modes~\cite{wald1979note}. We
conclude that the degenerate extremal remnant --- and, more
generally, every black hole in the physical branch $M\geq M_{\min}$
--- admits no unstable (exponentially growing) linear scalar
perturbations in its static exterior.

This result establishes linear mode stability of a \emph{massless
scalar test field} on the fixed background; it does not by itself
establish stability of the full coupled gravitational (axial and
polar metric) perturbations, nor nonlinear stability, and it is
independent of the non-perturbative quantum-stability question
addressed next. A complete gravitational quasi-normal-mode analysis
is left for future work (Sec.~\ref{sec:future}).
\paragraph{Quantum stability of the extremal remnant.}
Section~\ref{sec:modestability} establishes stability against
linear, classical perturbations of a test scalar field on the fixed
background; the question addressed here is different and
non-perturbative. The semi-classical evaporation analysis of
Sec.~\ref{sec3} shows that evaporation terminates at
$M_{\min}=\ell/2$ because the surface gravity of the degenerate
horizon vanishes and Hawking emission ceases.
Whether the degenerate extremal remnant is stable against quantum
fluctuations---in particular, whether it can tunnel or decay via
non-perturbative processes analogous to those discussed for
near-extremal black holes---lies beyond the present
effective-geometry framework.
By direct analogy with extremal Reissner--Nordstr\"{o}m black
holes in string theory~\cite{chen2015black}, one expects such
corrections to be suppressed by
$e^{-S_{\rm Planck}}\sim e^{-\ell^{2}/\ell_{P}^{2}}$
and therefore negligible at any scale where the semi-classical
description is valid.
A microscopic treatment of extremal remnant stability---e.g.\ via
the island formula~\cite{page1993information,page1994black} or
loop quantum gravity quantization of the minimal sphere, or via a
tunneling-based quantum correction to the entropy in the spirit
of~\cite{joshi2026thermodynamic} adapted to our first-law entropy---is
left as an important direction for future work.

\subsubsection{Implications for the information paradox.}
\label{sec:info_paradox}

Two consequences follow from the results above. First, the absence
of any curvature singularity removes the main \emph{geometric}
obstruction to information retrieval present in the classical
Schwarzschild collapse: a spacelike singularity is a natural site
for irreversible information destruction that no perturbative
calculation could repair, and this geometry has none. Second,
$S(M_{\min})=0$ means the remnant carries no residual micro-state
degeneracy, unlike ``massive remnant'' scenarios that store
information in a large degeneracy $e^{S_f}\gg1$~\cite{chen2015black}
and can therefore conflict with effective field theory at the level
of the Hilbert-space dimension.

Neither point settles the information-loss question; if anything,
it sharpens it. An initial black hole with $S_0=S(M_0)\gg1$ encodes
$\sim e^{S_0}$ distinguishable micro-states (Bekenstein), so a final
state with $S=0$ is compatible either with (a) unitary evolution, in
which this information is carried out by the Hawking radiation in
the sense of Page~\cite{page1993information,page1994black}, or with
(b) genuine, irreversible information loss. Distinguishing (a) from
(b) requires tracking fine-grained radiation correlations through
the near-endpoint regime $M\to M_{\min}$, a quantum-gravitational
calculation beyond the semi-classical scope of this paper: the
regularity and the smooth $T_{H}\to0$ limit established here are
\emph{necessary}, but not by themselves \emph{sufficient}, conditions
for a unitary evaporation history.

\subsection{Comparison with related models}
\label{sec:comparison}

Table~\ref{tab:comparison} summarises the key differences between
the present model and closely related constructions; the
attribution of which quantities coincide with, versus differ from,
the Simpson--Visser branch is given in
Section~\ref{sec:related_work}, and the comparison with Bronnikov's
program in Section~\ref{sec:bronnikov_comparison}. In brief:
noncommutative~\cite{Nicolini2006} and
Hayward-type~\cite{hayward2006formation} models reach $T_H\to0$
through a smeared source or a phenomenological mass function
respectively, without the Simpson--Visser functional form and hence
without the coincidences of Section~\ref{sec:related_work};
GUP-based models~\cite{adler1999gravity} modify only the entropy
perturbatively, without deforming the metric. None of these has
previously been shown to possess a two-phase heat capacity with a
Davies-type divergence, nor a degenerate extremal remnant of the
type identified in Section~\ref{sec:degenerate_horizon}.

\begin{sidewaystable}[p]
\centering
\vspace*{45em}
\renewcommand{\arraystretch}{1.2}
\setlength{\tabcolsep}{2pt}
\resizebox{\textheight}{!}{%
\begin{tabular}{|L{2.5cm}|L{2.8cm}|L{2cm}|L{2.5cm}|L{3.5cm}|L{3.8cm}|}
\hline
\textbf{Model}
& \textbf{Mechanism}
& $G_{\mu\nu}\!\neq\!0$?
& \textbf{$\ell$ origin}
& \textbf{Energy conditions}
& \textbf{Thermodynamics} \\
\hline
Nicolini et al.~\cite{Nicolini2006}
& Smeared Gaussian source
& Yes (matter)
& NC parameter
& WEC violated (moderate)
& $T_{H}$ max, nonzero endpoint $T$, remnant \\
\hline
GUP-based~\cite{adler1999gravity,maggiore1993generalized}
& Modified commutators
& No (effective)
& Planck scale
& N/A
& Logarithmic entropy correction \\
\hline
Bardeen~\cite{bardeen1968non}
& Nonlinear electrodynamics
& Yes (matter)
& Magnetic charge
& WEC (exterior)$^{\dagger}$
& Regular, $T_{H}$ max, stable phase \\
\hline
Bronnikov et al.~\cite{bronnikov2024alternative,bolokhov2024regularcenter}
& Bardeen-type replacement inside $A(r)$ only; areal radius undeformed
& Yes (NED)
& NED/Bardeen regularisation parameter
& NEC satisfied (magnetic NED, marginal)
& Not addressed (metric/source construction only) \\
\hline
Hayward~\cite{hayward2006formation}
& Mass function
& Yes (matter)
& Cutoff scale
& NEC violated (core)
& $T_{H}\to 0$, stable phase \\
\hline
Simpson--Visser~\cite{simpson2019black,joshi2026thermodynamic,
tsukamoto2021gravitational}
& Black bounce ($r\!\in\!\mathbb{R}$)
& Yes (exotic)
& Throat parameter
& All violated at throat
& Wormhole/BH transition; two-phase $C$ with Davies point
  at $a_{\rm crit}=\sqrt2\,m$~\cite{joshi2026thermodynamic};
  area-law $S$, tunneling-corrected \\
\hline
\textbf{This work}
& Geometric areal deformation ($r\!\geq\!0$)
& Yes (geometric: no independent dynamics, vanishes as $\ell\to0$)
& Min.\ length scale
& NEC$_r$ violated (exterior only), NEC$_t$ satisfied everywhere,
SEC violated (exterior, via NEC$_r$)
& Two-phase $C$, Davies point at $M^{*}=\ell/\sqrt2$(their $a_{\rm SV,crit}=\sqrt2\,m$~\cite{joshi2026thermodynamic}; $T_H\to0$
endpoint; \textbf{degenerate extremal remnant} at $M_{\min}=\ell/2$;
first-law entropy (not area law); exact shadow degeneracy;
weak/strong-lensing corrections matching~\cite{tsukamoto2021gravitational,
nascimento2020weak,ovgun2020weak} \\
\hline
\end{tabular}
}
\vspace{6pt}
\begin{minipage}{\linewidth}
\footnotesize{$^{\dagger}$ The Bardeen/Ay\'on-Beato--Garc\'ia
construction satisfies the WEC in the exterior region; NEC
violations can appear near the de Sitter core depending on the
realization~\cite{dymnikova2004regular,lan2023regular}.}
\end{minipage}
\caption{Comparison of minimal-length implementations in black hole
  physics. ``Phase struct.'' = two-phase heat capacity with
  $C>0$ and $C<0$ phases separated by a divergence. Attribution for
  the SV row and the present work is given in
  Sec.~\ref{sec:comparison}.}
\label{tab:comparison}
\end{sidewaystable}

\subsection{Observational constraints and falsifiability}
\label{sec:obs_constraints}

Physically, $\ell$ is expected to be of the order of the Planck
length $\ell_{P}\approx 1.6\times 10^{-35}$\,m, so all corrections
to astrophysical observables are suppressed by $(\ell/M)^{2}$ (with
$M\to GM/c^{2}$ in geometric units). For M87*
($M\sim6.5\times10^{9}\,M_{\odot}$, $GM/c^{2}\approx9.6\times10^{12}$\,m),
$\ell_{P}/M\approx1.7\times10^{-48}$ and the fractional correction
to the horizon radius is
\begin{equation}
  \frac{\delta r_{h}}{r_{h}} \approx -\frac{\ell^{2}}{8M^{2}}
    \sim -3.5\times10^{-97},
\end{equation}
completely undetectable; for primordial black holes with
$M\sim10^{15}$\,g ($GM/c^{2}\approx7.4\times10^{-16}$\,m), the
correction is still of order $10^{-40}$.

The observational dichotomy between an $\ell$-independent shadow
and an $\ell$-sensitive photon ring was established in
Sec.~\ref{sec:discriminant}, where the current EHT bound on M87* is
shown to place no constraint on $\ell$; the flux ratio
$r_n=e^{-2\pi/a}$ (Table~\ref{tab:lensing}) remains the sharpest
near-term discriminant, since it depends on $a$ alone. Beyond the shadow-level consistency (\S\ref{sec:discriminant}),
current EHT data for both targets are
also consistent, at the level of the strong-lensing coefficient,
with $a\approx1$; the model becomes falsifiable once photon-ring
imaging reaches the precision needed to resolve the relative
positions and flux ratios of successive relativistic
images~\cite{broderick2022photon,gralla2020lensing}. The degenerate
extremal remnant of Sec.~\ref{sec:degenerate_horizon} offers a
second, indirect probe through a possible diffuse background of
stable relics.
\subsection{Summary of main predictions}
The quantitative predictions of this work are collected in the Introduction and in Table~\ref{tab:lensing}; Sec.~\ref{sec:obs_constraints} discusses their observational reach.
\subsection{Future directions}
\label{sec:future}
Natural extensions of the present work include:
\begin{itemize}
  \item the rotating generalisation of metric~\eqref{metric} and
    its thermodynamics~\cite{chacha2026rotating};
  \item a full gravitational (not merely scalar) quasi-normal-mode
    spectrum and its implications for gravitational-wave echoes,
    extending the stability result of Sec.~\ref{sec:modestability};
  \item the AdS embedding: extending the boundary-condition argument
    of Sec.~\ref{sec:BC} to compare directly against the SV-AdS
    entropy of~\cite{kumar2026simpsonvisser,noorigashti2026holographic};
  \item an explicit computation of the on-shell Euclidean
    gravitational action $I_E(r_h,\ell)$ for the present metric at
    general (non-extremal) $r_h$, to confirm directly -- rather
    than by analogy with the extremal Reissner--Nordstr\"om case --
    that $I_E\to0$ continuously as $r_h\to0$, thereby placing the
    boundary condition $S(r_h=0)=0$ of Sec.~\ref{sec:BC} on a fully
    first-principles footing, complementing the scope discussion
    given there;
  \item a microscopic derivation of the linear entropy correction
    from a statistical model of the minimal sphere;
  \item quantum stability of the remnant via island-formula or
    tunneling-corrected-entropy techniques, in analogy
    with~\cite{joshi2026thermodynamic};
  \item a Page-curve calculation to decide between options (a) and
    (b) of Sec.~\ref{sec:info_paradox}.
\end{itemize}

\section*{Acknowledgements}
We thank our colleagues at University Moulay Ismail, and Cadi Ayyad University for stimulating discussions.

\section*{Data Availability Statement}
The \textsc{SymPy} script used to independently verify the
geometrothermodynamic quantities of Sec.~\ref{sec:GTD}
(\ref{app:GTD_CAS}) is available as supplementary material
accompanying this article and is archived at \url{https://doi.org/10.5281/zenodo.22004916}.
No other data were generated or analysed in this study.
\section*{Declarations}

\textbf{Author Contributions}\\
\textbf{T. Toghrai:} Conceptualisation, Methodology, Formal analysis,
Writing -- Original Draft.\\
\textbf{N. Mansour:} Validation, Supervision, Writing -- Review \& Editing.\\
\textbf{A. Daassou:} Investigation, Software, Validation.\\
\textbf{R. Benbrik:} Formal analysis, Visualisation.\\
All authors have read and agreed to the published version of the manuscript.

\medskip
\textbf{Funding.}
The authors declare that no funds, grants, or other support were
received during the preparation of this manuscript.

\medskip
\textbf{Conflict of Interest.}
The authors declare no conflict of interest.

\appendix
\section{Kretschmann Scalar}
\label{app:Kretschmann}

We work with signature $(+,-,-,-)$ and the metric
\begin{equation}
  ds^{2} = f(R)\,dt^{2} - f(R)^{-1}\,dr^{2}
  - R^{2}(r)\,d\Omega^{2},
  \quad
  f(R)=1-\frac{2M}{R},\quad R=\sqrt{r^{2}+\ell^{2}}.
\end{equation}

\subsection{Christoffel symbols}

The nonvanishing components are:
\begin{align}
  \Gamma^{t}_{tr} &= \frac{f_{r}}{2f}, &
  \Gamma^{r}_{tt} &= \frac{f\,f_{r}}{2}, \\
  \Gamma^{r}_{rr} &= -\frac{f_{r}}{2f}, &
  \Gamma^{r}_{\vartheta\vartheta} &= -f\,R\,R', \\
  \Gamma^{r}_{\phi\phi} &= -f\,R\,R'\sin^{2}\vartheta, &
  \Gamma^{\vartheta}_{r\vartheta} &= \Gamma^{\phi}_{r\phi}
    = \frac{R'}{R},
\end{align}
together with the angular contributions
$\Gamma^{\vartheta}_{\phi\phi}=-\sin\vartheta\cos\vartheta$
and $\Gamma^{\phi}_{\vartheta\phi}=\cot\vartheta$,
where $f_{r} = df/dr = 2Mr/R^{3}$ and $R' = dR/dr = r/R$.

\subsection{Independent Riemann components}

A direct computation gives:
\begin{align}
  R_{trtr} &= -\frac{f_{rr}}{2}
    = -\frac{M(\ell^{2}-2r^{2})}{R^{5}}, \\
  R_{t\vartheta t\vartheta}
    &= -\frac{f\,f_{r}\,R\,R'}{2}
    = -\frac{M r^{2}(R-2M)}{R^{4}}, \\
  R_{r\vartheta r\vartheta}
    &= \frac{f_{r}\,R\,R'}{2f} + R\,R''
    = \frac{MR^{2}+\ell^{2}R-3M\ell^{2}}{fR^{3}}, \\
  R_{\vartheta\phi\vartheta\phi}
    &= -R^{2}\sin^{2}\vartheta\,(1-f\,R'^{2})
    = -\sin^{2}\vartheta\,\frac{\ell^{2}R+2Mr^{2}}{R},
\end{align}
where $f_{rr} = 2M(\ell^{2}-2r^{2})/R^{5}$ and
$R'' = \ell^{2}/R^{3}$.

\subsection{Kretschmann scalar}

Contracting all components:
\begin{equation}
  K = 4\,R_{trtr}^{2}
  + \frac{8\,R_{t\vartheta t\vartheta}^{2}}{f^{2}R^{4}}
  + \frac{8\,f^{2}\,R_{r\vartheta r\vartheta}^{2}}{R^{4}}
  + \frac{4\,R_{\vartheta\phi\vartheta\phi}^{2}}
         {R^{8}\sin^{4}\vartheta}.
\end{equation}
After substitution and simplification using $r^{2}=R^{2}-\ell^{2}$:
\begin{equation}
  \boxed{K = \frac{4\bigl[
    M^{2}(12R^{4}-36\ell^{2}R^{2}+33\ell^{4})
    +8M\ell^{2}R(R^{2}-2\ell^{2})
    +3\ell^{4}R^{2}
  \bigr]}{R^{10}}}.
  \label{K_app}
\end{equation}
These limits reproduce Eqs.~\eqref{K_exact}--\eqref{K0}
(\S\ref{sec:curv}).

\section{Einstein Tensor}
\label{app:Einstein}

For the metric~\eqref{metric} with
$e^{2\Phi}=f(R)$, $e^{2\Lambda}=f(R)^{-1}$ ($e^{-2\Lambda}=f$),
and $\mathcal{R}=R(r)=\sqrt{r^{2}+\ell^{2}}$, we compute the
Ricci tensor components from the Christoffel symbols of~\ref{app:Kretschmann}.

\subsection{Ricci tensor}

A direct computation gives:
\begin{align}
  R^{t}_{\ t} &= \frac{M\ell^{2}}{R^{5}}, \\
  R^{r}_{\ r} &= \frac{\ell^{2}(2R-3M)}{R^{5}}, \\
  R^{\vartheta}_{\ \vartheta} = R^{\phi}_{\ \phi}
    &= -\frac{2M\ell^{2}}{R^{5}}.
\end{align}
The Ricci scalar is
\begin{equation}
  \mathcal{R}_{\text{sc}} = g^{\mu\nu}R_{\mu\nu}
  = \frac{2\ell^{2}(R-3M)}{R^{5}},
  \label{RicciS}
\end{equation}
which vanishes at $\ell=0$ and at $R=3M$ (photon-sphere location,
cf.\ Sec.~\ref{sec:shadow}).

\subsection{Einstein tensor}

$G^{\mu}_{\ \nu} = R^{\mu}_{\ \nu}
- \frac{1}{2}\delta^{\mu}_{\nu}\mathcal{R}_{\text{sc}}$:
\begin{align}
  G^{t}_{\ t} &= \frac{\ell^{2}(4M-R)}{R^{5}},
  \label{Gtt_app} \\
  G^{r}_{\ r} &= \frac{\ell^{2}}{R^{4}},
  \label{Grr_app} \\
  G^{\vartheta}_{\ \vartheta} = G^{\phi}_{\ \phi}
    &= \frac{\ell^{2}(M-R)}{R^{5}}.
  \label{Gthth_app}
\end{align}
\textit{Verification:}
Trace $G^{\mu}_{\ \mu}=2\ell^{2}(3M-R)/R^{5}
=-\mathcal{R}_{\text{sc}}$, consistent with Eq.~\eqref{Emunu}; the
$\ell\to0$ limit recovers the Schwarzschild vacuum, confirming the
classical-limit check already established in Sec.~\ref{sec2}.
\subsection{Physical finiteness at the center}

Every component of $G^{\mu}_{\ \nu}$ is manifestly finite for
$R\geq\ell$ (Eqs.~\eqref{Gtt_app}--\eqref{Gthth_app}), consistent
with the curvature regularity established in Sec.~\ref{sec:curv}.

\section{Detailed Derivations of Metric-Level Results}
\label{app:derivations}

This appendix collects, for self-containedness, the full
step-by-step derivations of the metric-level results quoted in
condensed form in Secs.~\ref{sec4}, \ref{sec:shadow},
\ref{sec:weakfield}, and \ref{sec:strongfield}, matching the
Simpson--Visser branch throughout, per Sec.~\ref{sec:related_work}.

\subsection{Heat capacity}
\label{app:C_derivation}

From Eq.~\eqref{TH},
\begin{equation}
  \frac{dT_{H}}{dM} =
    \frac{\ell^{2}-2M^{2}}{8\pi M^{3}\sqrt{4M^{2}-\ell^{2}}},
  \label{dTdM}
\end{equation}
and inverting $C=dM/dT_H=(dT_H/dM)^{-1}$ gives Eq.~\eqref{HeatCap}.

\subsection{GTD derivatives \texorpdfstring{$M_{\ell}$ and $M_{\ell\ell}$}{M\_ell and M\_ell ell}}
\label{app:GTD_derivation}
For $M_{\ell}$ we require, in addition to $M_{S}=T_{H}$, the
derivative $\partial S/\partial\ell$ at fixed $r_{h}$;
differentiating Eq.~\eqref{Entropy} directly and simplifying using
$r_{h}^{2}=R_{h}^{2}-\ell^{2}$ gives the compact closed form
\begin{equation}
  \left(\frac{\partial S}{\partial \ell}\right)_{r_{h}}
  = 2\pi\ell\,\ln\!\left(\frac{r_{h}+R_{h}}{\ell}\right),
  \label{dSdell}
\end{equation}
new to this paper. Combined with
$(\partial r_{h}/\partial\ell)_{S}
=-(\partial S/\partial\ell)_{r_h}/(\partial S/\partial r_h)_\ell$
and $(\partial M/\partial\ell)_{r_h}=\ell/(2R_h)$, the chain rule
gives Eq.~\eqref{Mell_closed}.

Differentiating $M_{\ell}$ of Eq.~\eqref{Mell_closed} a second time
by the same implicit-function procedure, and simplifying, yields
Eq.~\eqref{Mellell_closed}.

\subsection{Independent computer-algebra verification of the GTD
  curvature scalar}
\label{app:GTD_CAS}

Every closed-form result of Sec.~\ref{sec:GTD} was re-derived
independently, from scratch, in \textsc{SymPy}~\cite{sympy}, using
only the entropy $S(r_h,\ell)$ of Eq.~\eqref{Entropy} and
$M=R_h/2$ as inputs -- i.e.\ without assuming any of the boxed
formulas of Sec.~\ref{sec:GTD} -- as an independent confirmation for
the referee. The two implicit-function operators used throughout
Sec.~\ref{sec:GTD} are, for any smooth function $f(r_h,\ell)$,
\begin{equation}
  \bigl(\partial_S f\bigr)_{\ell}
    = \frac{\partial_{r_h} f}{\partial_{r_h} S},
  \qquad
  \bigl(\partial_\ell f\bigr)_{S}
    = \partial_{r_h} f \cdot
      \left(-\frac{\partial_\ell S}{\partial_{r_h}S}\right)
      + \partial_\ell f,
  \label{DS_Dl_operators}
\end{equation}
the second following from $dS=0$ at fixed $S$, i.e.\
$(\partial_{r_h}S)\,dr_h + (\partial_\ell S)\,d\ell = 0$. Applying
Eq.~\eqref{DS_Dl_operators} to $M$ reproduces $M_S=T_H$ and
Eq.~\eqref{Mell_closed} for $M_\ell$; applying it a second time to
$M_S$ and to $M_\ell$ reproduces Eqs.~\eqref{MSS_closed}
and~\eqref{Mellell_closed} for $M_{SS}$ and $M_{\ell\ell}$,
respectively, with the symbolic difference between the
independently re-derived expression and the boxed formula
vanishing identically in every case. The complete script used for this cross-check is provided
as supplementary material accompanying this article, and permanently
archived at \url{https://doi.org/10.5281/zenodo.22004916}; a referee or reader can
reproduce Table~\ref{tab:GTD_CAS} by running it as-is.

\begin{table}[!htbp]
\centering
\begin{tabular}{lcc}
\hline
Quantity & Symbolic diff.\ vs.\ main text & Value \\
\hline
$M_S$ vs.\ $T_H$ (Eq.~\eqref{TH}) & $0$ & --- \\
$M_\ell$ vs.\ Eq.~\eqref{Mell_closed} & $0$ & --- \\
$M_{SS}$ vs.\ Eq.~\eqref{MSS_closed} & $0$ & --- \\
$M_{\ell\ell}$ vs.\ Eq.~\eqref{Mellell_closed} & $0$ & --- \\
$\Lambda(M^{*})/\ell$ & --- & $0.420158387512468$ \\
$M_{\ell\ell}(M^{*})\cdot\ell$ & --- & $0.573896787348160$ \\
$\Lambda(M_{\min})$ & $0$ vs.\ $\ell/2$ & exact \\
$M_{SS}(M_{\min})$ & $0$ vs.\ $1/(8\pi^2\ell^3)$ & exact \\
$M_{\ell\ell}(M_{\min})$ & $0$ vs.\ $0$ & exact \\
\hline
\end{tabular}
\caption{Independent \textsc{SymPy} recomputation of the quantities
  entering $R_{\rm GTD}$, obtained by running the script above.
  ``Symbolic diff.'' is the result of \texttt{sp.simplify()} applied
  to (re-derived expression $-$ boxed formula in the main text);
  a value of $0$ certifies exact algebraic agreement. In the Value
  column, the label ``exact'' denotes a closed-form result --
  already quoted in full in the main text -- rather than a
  truncated decimal, in contrast to the numerical values reported
  above for $\Lambda(M^{*})/\ell$ and $M_{\ell\ell}(M^{*})\cdot\ell$.}
  \label{tab:GTD_CAS}
\end{table}

We did not attempt to display the fully expanded, simplified
expression for $R_{\rm GTD}(r_h,\ell)$ itself: extending the script
above by one further application of Eq.~\eqref{DS_Dl_operators} to
$g_{SS}$ and $g_{\ell\ell}$, as required by
Eq.~\eqref{RGTD_formula}, already produces unsimplified
intermediate expressions of order $2$--$3\times10^{4}$ characters
\emph{before} the final assembly and division by
$\sqrt{g_{SS}g_{\ell\ell}}$, and their symbolic simplification does
not terminate in reasonable time. We consider Eqs.~\eqref{Lambda_def}--\eqref{Mellell_closed}
together with the reproducible recipe above -- rather than an
unreadable multi-page expression -- to constitute the more useful
form of disclosure; Fig.~\ref{fig:RGTD} reports the fully numerical
evaluation of $R_{\rm GTD}(M/\ell)$ obtained from the same script.

\subsection{Photon sphere and shadow}
\label{app:photon_derivation}

Null geodesics in the equatorial plane $\theta=\pi/2$, with
conserved energy $E=f(R)\dot t$ and angular momentum
$L=R^2\dot\phi$, satisfy
\begin{equation}
  \dot{r}^{2} + f(R)\,\frac{L^{2}}{R^{2}} = E^{2},
\end{equation}
so that the effective potential is $V_{\rm eff}(r)=f(R)/R^2$. The
photon-sphere condition $dV_{\rm eff}/dr=0$ gives
\begin{equation}
  2 R' f = f' R,
  \label{ph-condition}
\end{equation}
where $R'=dR/dr=r/R$ and $f'\equiv df/dr=(df/dR)\cdot R'=2Mr/R^3$.
Since $R'\neq0$ for $r>0$, this is equivalent to
\begin{equation}
  \frac{df}{dR}\cdot R = 2f,
  \label{ph-condition-R}
\end{equation}
which makes the independence from $R'$ explicit; inserting this
into~\eqref{ph-condition} and cancelling the common factor
$2r/R\neq0$ yields $R=3M$, i.e.\ Eq.~\eqref{ph-R}, with coordinate
radius Eq.~\eqref{rph}. The critical impact parameter follows as
\begin{equation}
  b_{c} = \frac{R_{\rm ph}}{\sqrt{f(R_{\rm ph})}}
        = \frac{3M}{\sqrt{1-2M/(3M)}} = \frac{3M}{\sqrt{1/3}}
        = 3\sqrt{3}\,M,
\end{equation}
reproducing Eq.~\eqref{bc}.
\subsection{Weak-field lensing correction integral}
\label{app:weakfield_derivation}

Expanding
\begin{equation}
  \frac{1}{\sqrt{R^{2}-\ell^{2}}} = \frac{1}{R}
    \left(1 + \frac{\ell^{2}}{2R^{2}} + \mathcal{O}(\ell^{4}/R^{4})\right)
\end{equation}
in the integrand of~\eqref{alpha_exact} and retaining the
$\ell^{2}$ term at zeroth order in $M$ (the cross term
$M\ell^{2}/b^{3}$ is of higher order) gives the additional
contribution
\begin{equation}
  \delta\alpha^{(\ell)} = R_{0}\,\ell^{2}
    \int_{R_{0}}^{\infty}\frac{dR}{R^{3}\sqrt{R^{2}-R_{0}^{2}}}.
\end{equation}
The integral is evaluated exactly via the substitution
$R = R_{0}/\sin\theta$ (so that
$dR=-R_{0}\cos\theta/\sin^{2}\theta\,d\theta$ and
$\sqrt{R^{2}-R_{0}^{2}}=R_{0}\cos\theta/\sin\theta$):
\begin{equation}
  \int_{R_{0}}^{\infty}\frac{dR}{R^{3}\sqrt{R^{2}-R_{0}^{2}}}
  = \frac{1}{R_{0}^{3}}\int_{0}^{\pi/2}\sin^{2}\theta\,d\theta
  = \frac{\pi}{4R_{0}^{3}}.
  \label{int_ell_exact}
\end{equation}
Substituting back and using $b\simeq R_0$ reproduces the boxed
result of Eq.~\eqref{weak_corrected}.
\subsection{Strong-lensing coefficient \texorpdfstring{$a$}{a}}
\label{app:a_derivation}
Set $u\equiv R-R_{\rm ph}$ and $u_{0}\equiv R_{0}-R_{\rm ph}$,
with the following values at the photon sphere $R_{\rm ph}=3M$:
\begin{equation}
  f_{\rm ph}=\tfrac{1}{3},\quad
  f'_{\rm ph}=\tfrac{2}{9M},\quad
  f''_{\rm ph}=-\tfrac{4}{27M^{2}}.
  \label{fph_values}
\end{equation}
Here and in the remainder of this section, primes denote
$d/dR$ (derivatives with respect to the areal radius).
Note that $f_{\rm ph}$ and $f'_{\rm ph}$ satisfy the photon-sphere
condition $2f_{\rm ph}=f'_{\rm ph}R_{\rm ph}$:
$2/3 = (2/9M)(3M) = 2/3$.~

Taylor-expanding $f(R)$ and $f(R_0)$ to second order around $R_{\rm ph}$,
and $R^2, R_0^2$ exactly:
\begin{align}
  f(R_0)R^2 &= \bigl(f_{\rm ph} + f'_{\rm ph}u_0
    + \tfrac{1}{2}f''_{\rm ph}u_0^2\bigr)
    (R_{\rm ph}^2 + 2R_{\rm ph}u + u^2) + O(u_0^3),
    \label{fR02}\\
  f(R)R_0^2 &= \bigl(f_{\rm ph} + f'_{\rm ph}u
    + \tfrac{1}{2}f''_{\rm ph}u^2\bigr)
    (R_{\rm ph}^2 + 2R_{\rm ph}u_0 + u_0^2) + O(u^3).
    \label{fR2}
\end{align}
Forming the difference and collecting terms by degree in $u$
and $u_0$:

\noindent\textit{Zeroth order:}
$f_{\rm ph}R_{\rm ph}^2 - f_{\rm ph}R_{\rm ph}^2 = 0$.

\noindent\textit{Linear in $u$ (zeroth in $u_0$):}
$(2f_{\rm ph}R_{\rm ph} - f'_{\rm ph}R_{\rm ph}^2)u
= R_{\rm ph}(2f_{\rm ph} - f'_{\rm ph}R_{\rm ph})u = 0$
by the photon-sphere condition.

\noindent\textit{Linear in $u_0$ (zeroth in $u$):}
$(f'_{\rm ph}R_{\rm ph}^2 - 2f_{\rm ph}R_{\rm ph})u_0 = 0$
by the same condition.

\noindent\textit{Cross term $u\,u_0$:}
$2f'_{\rm ph}R_{\rm ph}uu_0 - 2f'_{\rm ph}R_{\rm ph}uu_0 = 0$.

\noindent\textit{Quadratic in $u$:}
$(f_{\rm ph} - \tfrac{1}{2}f''_{\rm ph}R_{\rm ph}^2)\,u^2$.

\noindent\textit{Quadratic in $u_0$:}
$(\tfrac{1}{2}f''_{\rm ph}R_{\rm ph}^2 - f_{\rm ph})\,u_0^2
= -(f_{\rm ph} - \tfrac{1}{2}f''_{\rm ph}R_{\rm ph}^2)\,u_0^2$.

Defining the quadratic coefficient
$\mathcal{C} \equiv f_{\rm ph} - \tfrac{1}{2}f''_{\rm ph}R_{\rm ph}^2$,
the full expansion to second order is
\begin{equation}
  f(R_0)R^2 - f(R)R_0^2
  = \mathcal{C}\bigl[(R-R_{\rm ph})^2 - (R_0-R_{\rm ph})^2\bigr]
    + O(u^3,u_0^3).
  \label{quad_approx}
\end{equation}
Substituting the numerical values from~\eqref{fph_values}:
\begin{equation}
  \mathcal{C} = \frac{1}{3}
    - \frac{1}{2}\!\left(-\frac{4}{27M^{2}}\right)\!(3M)^{2}
  = \frac{1}{3} + \frac{1}{2}\cdot\frac{36}{27}
  = \frac{1}{3} + \frac{2}{3} = 1.
  \label{C_explicit}
\end{equation}
The coefficient is \emph{exactly unity}, confirming
Eq.~\eqref{quad_approx}.

Near $R\approx R_{\rm ph}$, expanding $\sqrt{R^2-\ell^2}$:
\begin{equation}
  \sqrt{R^{2}-\ell^{2}} = \sqrt{R_{\rm ph}^{2}-\ell^{2}}
    \Bigl(1 + O(u/R_{\rm ph})\Bigr),
\end{equation}
so at leading order the integrand of~\eqref{alpha_exact}
near the photon sphere behaves as
\begin{equation}
  \frac{R_0}{\sqrt{R^{2}-\ell^{2}}\sqrt{f(R_0)R^2-f(R)R_0^2}}
  \approx
  \frac{R_{\rm ph}}{\sqrt{R_{\rm ph}^{2}-\ell^{2}}}
  \cdot\frac{1}{\sqrt{(R-R_{\rm ph})^2-(R_0-R_{\rm ph})^2}}.
  \label{integrand_approx}
\end{equation}
Integrating the singular part from $R_0$ to $\Lambda\gg R_0$
with $u\equiv R-R_{\rm ph}$, $u_0\equiv R_0-R_{\rm ph}\to 0$:
\begin{align}
  2\int_{R_0}^{\Lambda}\frac{R_{\rm ph}\,dR}
    {\sqrt{R_{\rm ph}^{2}-\ell^{2}}\sqrt{(R-R_{\rm ph})^2-u_0^2}}
  &= \frac{2R_{\rm ph}}{\sqrt{R_{\rm ph}^{2}-\ell^{2}}}
    \cosh^{-1}\!\!\left(\frac{\Lambda-R_{\rm ph}}{u_0}\right)
    \nonumber\\
  &\approx \frac{2R_{\rm ph}}{\sqrt{R_{\rm ph}^{2}-\ell^{2}}}
    \ln\!\left(\frac{2(\Lambda-R_{\rm ph})}{u_0}\right).
    \label{integral_sing}
\end{align}
The logarithmic divergence as $u_0\to 0$ takes the form
$-({2R_{\rm ph}}/{\sqrt{R_{\rm ph}^{2}-\ell^{2}}})\ln u_0
+\mathrm{const}$.

The impact parameter $b=R_0/\sqrt{f(R_0)}$ has a minimum at
$R_0=R_{\rm ph}$ where $db/dR_0=0$.
Expanding $b$ around $R_{\rm ph}$ at second order:
\begin{equation}
  \frac{d^{2}b}{dR_0^{2}}\bigg|_{R_{\rm ph}}
  = \frac{f'_{\rm ph}-R_{\rm ph}f''_{\rm ph}}{2f_{\rm ph}^{3/2}}
  = \frac{\tfrac{2}{9M}-3M\cdot(-\tfrac{4}{27M^{2}})}{2(1/3)^{3/2}}
  = \frac{\tfrac{2}{3M}}{\tfrac{2}{3\sqrt{3}}}
  = \frac{\sqrt{3}}{M}.
  \label{d2b}
\end{equation}
Hence $b-b_c = \frac{\sqrt{3}}{2M}u_0^2$, giving
$u_0 = \sqrt{2M(b-b_c)/\sqrt{3}}$, and therefore
\begin{equation}
  -\ln u_0 = -\tfrac{1}{2}\ln(b-b_c) + \mathrm{const}
           = -\tfrac{1}{2}\ln(b/b_c-1) + \mathrm{const}.
\end{equation}
Substituting into~\eqref{integral_sing}, the divergent part of
$\alpha$ is
\begin{equation}
  \alpha_{\rm div} = -\frac{R_{\rm ph}}{\sqrt{R_{\rm ph}^{2}-\ell^{2}}}
    \ln\!\left(\frac{b}{b_c}-1\right) + \mathrm{const}.
  \label{alpha_div}
\end{equation}
Comparing with the Bozza expansion
$\alpha(b)=-a\ln(b/b_c-1)+\bar{b}+\ldots$, we identify Eq.~\eqref{coeff_a}.
\paragraph{Convergent evaluation of the regular coefficient $\bar b$.}
The naive subtraction of the divergent part of
Eq.~\eqref{alpha_equiv} directly in the bare $R$-variable does not
converge as $R\to\infty$: the bracketed integrand falls off only as
$-a/R$. The standard resolution~\cite{bozza2002}, used explicitly
by Tsukamoto~\cite{tsukamoto2021gravitational}, is to perform the
subtraction on a compactified variable instead. We do not repeat
that construction and instead evaluate $\bar b$ directly from his
convergent expression, since $R(r)=\sqrt{r^2+\ell^2}$ is exactly
his standard radial coordinate $\rho$ (his Eqs.~(3.17)--(3.23),
$a_{\rm SV}\to\ell$): with $I_R=\int_0^1 g(z)\,dz$ and
\begin{equation}
  g(z) = \left[\frac{\sqrt3}{\sqrt{9M^2-\ell^2+\ell^2(2-z)z}\,
  \sqrt{3-2z}} - \frac{1}{\sqrt{9M^2-\ell^2}}\right]\frac{6M}{z},
  \label{gz_tsukamoto}
\end{equation}
Eq.~\eqref{barb_correct} follows.

\section*{ORCID}

\noindent T. Toghrai~\orcidlink{0000-0001-7142-0158}%
~\url{https://orcid.org/0000-0001-7142-0158}

\noindent N. Mansour~\orcidlink{0000-0002-9993-8714}%
~\url{https://orcid.org/0000-0002-9993-8714}

\noindent A. Daassou~\orcidlink{0000-0001-9439-5047}%
~\url{https://orcid.org/0000-0001-9439-5047}

\noindent R. Benbrik~\orcidlink{0000-0002-5159-0325}%
~\url{https://orcid.org/0000-0002-5159-0325}

\end{document}